\documentclass[%
 reprint,
superscriptaddress, 
 amsmath,amssymb,
 aps,
]{revtex4-1}

\usepackage[utf8]{inputenc}

\usepackage[T1]{fontenc}
\usepackage{multirow}
\usepackage{csquotes}

\usepackage{graphicx}
\usepackage{dcolumn}
\usepackage{bm}
\usepackage{float}
\usepackage[dvipsnames]{xcolor}
\colorlet{green1}{green!40!black!100!}
\newcommand{\vect}[1]{\boldsymbol{#1}}
\usepackage{hyperref}

\begin{document}

\preprint{APS/123-QED}

\title{Interactions between polyelectrolytes mediated by ordering and orientation of multivalent non-spherical ions in salt solutions}

\author{Hossein Vahid}
\affiliation{Department of Applied Physics, Aalto University, P.O. Box 11000, FI-00076 Aalto, Finland}
\affiliation{Department of Chemistry and Materials Science, Aalto University, P.O. Box 16100, FI-00076 Aalto, Finland}
\affiliation{Academy of Finland Center of Excellence in Life-Inspired Hybrid Materials (LIBER), Aalto University, P.O. Box 16100, FI-00076 Aalto, Finland}

\author{Alberto Scacchi}
\affiliation{Department of Applied Physics, Aalto University, P.O. Box 11000, FI-00076 Aalto, Finland}
\affiliation{Department of Chemistry and Materials Science, Aalto University, P.O. Box 16100, FI-00076 Aalto, Finland}
\affiliation{Academy of Finland Center of Excellence in Life-Inspired Hybrid Materials (LIBER), Aalto University, P.O. Box 16100, FI-00076 Aalto, Finland}

\author{Maria Sammalkorpi}
\affiliation{Department of Chemistry and Materials Science, Aalto University, P.O. Box 16100, FI-00076 Aalto, Finland}
\affiliation{Academy of Finland Center of Excellence in Life-Inspired Hybrid Materials (LIBER), Aalto University, P.O. Box 16100, FI-00076 Aalto, Finland}
\affiliation{Department of Bioproducts and Biosystems, Aalto University, P.O. Box 16100, FI-00076 Aalto, Finland}

\author{Tapio Ala-Nissila}
\email{tapio.ala-nissila@aalto.fi}
\affiliation{Department of Applied Physics, Aalto University, P.O. Box 11000, FI-00076 Aalto, Finland}
\affiliation{Quantum Technology Finland Center of Excellence, Department of Applied Physics, Aalto University, P.O. Box 11000, FI-00076 Aalto, Finland}
\affiliation{Interdisciplinary Centre for Mathematical Modelling and Department of Mathematical Sciences, Loughborough University, Loughborough, Leicestershire LE11 3TU, UK}

\date{\today}


\begin{abstract}
{
Multivalent ions in solutions with polyelectrolytes (PE) induce electrostatic correlations that can drastically change ion distributions around the PEs and their mutual interactions. 
Using coarse-grained molecular dynamics simulations, we show how in addition to valency, ion shape and concentration can be harnessed as tools to control like-charged PE-PE interactions. We 
demonstrate a correlation between the orientational ordering of aspherical ions and how they mediate the effective PE-PE attraction induced by multivalency. The interaction type, strength and range can thus be externally controlled in ionic solutions. Our results can be used as generic guidelines to tune the self-assembly of like-charged polyelectrolytes by variation of the characteristics of the ions.
}
\end{abstract}

\maketitle

Electrostatic interactions between charged molecules and ions in solution are ubiquitous in colloidal, soft, and biological systems~\cite{israelachvili2011}.
Systems such as some polyelectrolytes (PEs), synthetic and biopolymers, DNA~\cite{gelbart2000, deserno2001}, nanotubes in phospholipids~\cite{brazhnik2005}, actin filaments~\cite{tang1997, angelini2003}, microtubules~\cite{needleman2004}, viruses~\cite{butler2003,buck2018}, and even bacteria~\cite{shi2018}, can often be approximated by charged cylinders immersed in an electrolyte solution consisting of a solvent and mobile ions~\cite{wong2010}.
Understanding the ion distribution in such systems is paramount since solution-mediated interactions are greatly affected by the ionic environment, especially due to electrostatic screening effects~\cite{matthew1985} and ion redistribution~\cite{ kuzovkov2014, adibnia2019}.
Both the nature and concentration of ions play a significant role, whence ion valency is an important handle for tuning the properties of macroions~\cite{grosberg2002, levin2002, wong2006, iwaki2007, dukhin2010}.

Many chemically specific ions, such as, e.g., diamine, spermine, and spermidine, exhibit elongated, cylindrical shapes and are multivalent~\cite{gosule1976, needleman2004}. Some anions in battery electrolytes are non-spherical, which influences ion transport and conductivity in solution \cite{Mauger2018,Marcinek2015}. Ionic liquids are typically composed of highly non-spherical ions, which influences their cohesion energy and maintains their liquid character, but also influences ionic transport \cite{Lewandowski2009,Marcinek2015}.  
Consequently, ion specificity is paramount in controlling interactions between charged macromolecules.
 
For modeling purposes, traditional mean-field approaches such as Poisson-Boltzmann (PB) theory treat mobile ions as point charges in the weak-coupling regime. The standard PB theory cannot describe general chemically specific ions~\cite{Ren2012,grochowski2008}. 
However, successful models incorporating ion size properties exist, such as those in Refs. \cite{borukhov1997,lopez2011,zhou2011,colla2017, batys2017}. In the case of like-charged PE-PE interactions, the PB theory always predicts repulsion. To this end, the soft-potential-enhanced-PB (SPB) theory has been shown to accurately predict ion distributions around PEs~\cite{vahid2022} for ion sizes up to the PE radius, and like-charged PE-PE repulsion for small monovalent ions such as Na$^+$ and Cl$^-$ \cite{yang2022}. In Ref.
\cite{jimenez2006}, however, like-charged PE-PE attraction was reported for large monovalent ions and high salt concentration, and attributed to short-range charge correlations beyond the PB theory.

For multivalent ions, charge-charge correlations naturally appear and cause charge reversal (see, e.g., Refs.~\onlinecite{jonsson2004,matsarskaia2020, antila2017}) in the strong-coupling regime even with point-like ions~\cite{moreira2000, deserno2000, grosberg2002, messina2009, buyukdagli2020, buyukdagli2019, buyukdagli2017, buyukdagli2014, hatlo2010_2,buyukdagli2012}. 
Both valency and ion size have been considered in the context of classical density functional theory~\cite{evans1979nature}; see, e.g., the recent advances on electric double layers~\cite{gonzalez2018, cats2021, cats2022}. There also exist Monte Carlo studies on dumbbell-like (two separated point charges), yet volumeless, ions focused on counterion-mediated interactions between charged plates~\cite{kim2008, may2008, hatlo2010, grime2010, bohinc2012, bossa2018} or cylinders~\cite{cha2018}. 
References~\onlinecite{hunter2001,maset2009,bohinc2012,bohinc2016,cha2018} have suggested that a bridging mechanism is responsible for the attraction between like-charged surfaces. Nevertheless, to our knowledge none of these approaches have simultaneously considered both correlations and steric effects of aspherical multivalent ions.

In this Letter, we extend the ion-mediated interaction scenario in the case of spheroidal multivalent ions. Using coarse-grained (CG) molecular dynamics (MD) simulations, we focus on systems composed of single and double rod-like PEs.
We first investigate the condensation and orientation response for different ion specificities around a single PE. We then address the order-mediated interactions between two like-charged PEs and underpin the effect of ion orientation, valency, and shape.
Tuning the balance between attractive and repulsive forces allows for controlling the thermodynamic properties of assembled systems, their stability, and their response to external conditions~\cite{gosule1976,  pelta1996, raspaud1998, butler2003, needleman2004,huang2017}. We show how this control can be achieved by adjusting ion valency, shape, and salt.

\begin{figure}[b!]
\centering
  \includegraphics[width=1\linewidth]{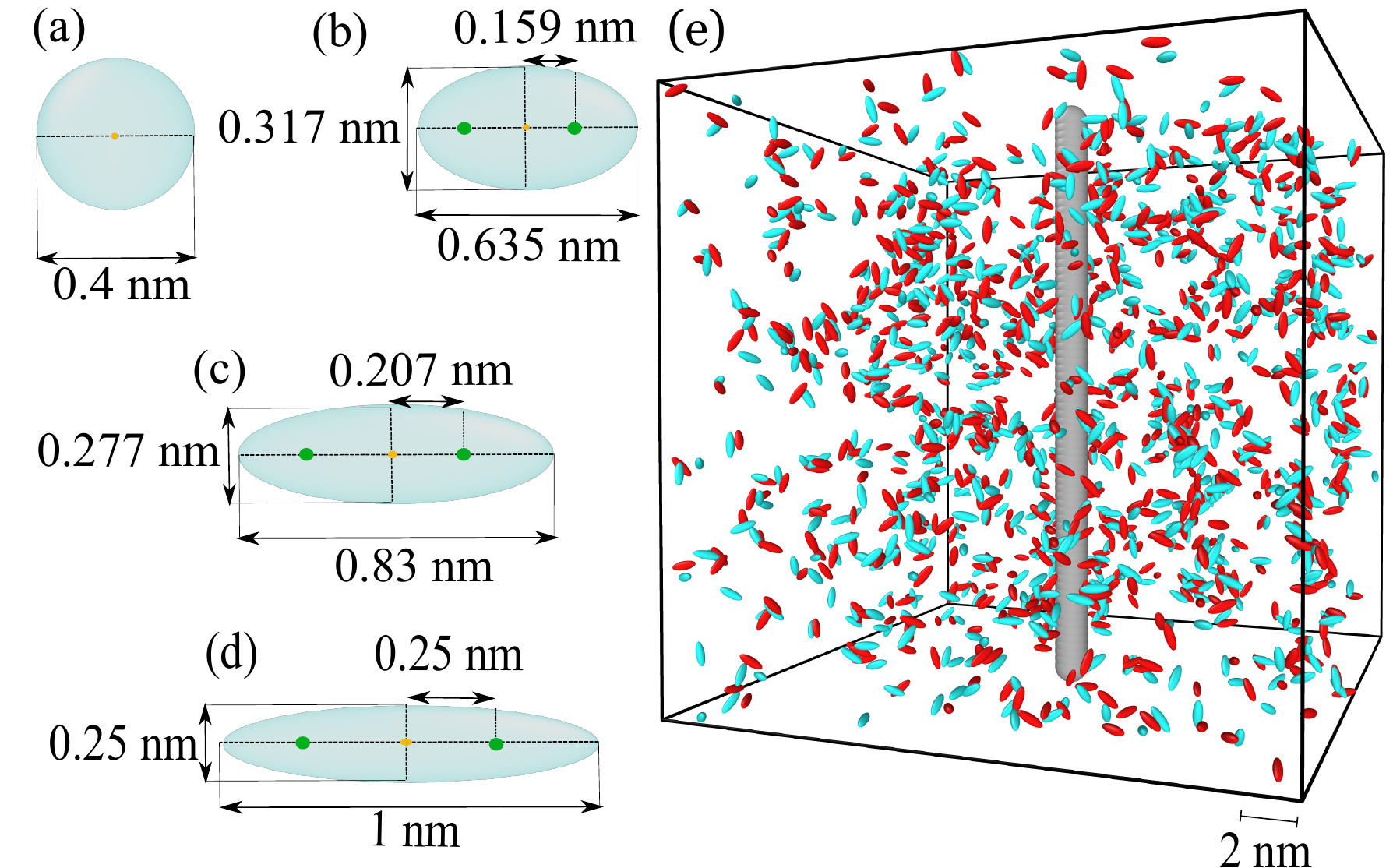}
 \caption{Ions with (a) $A^{\rm c}=1$, (b) $A^{\rm c}=2$, (c) $A^{\rm c}=3$, and (d) $A^{\rm c}=4$. The charges are separated from the center by $\sigma_{\rm maj}/4$ (green dots). (e) Snapshot of the simulation box of size (20 nm)$^3$ with periodic boundary conditions containing one PE (grey), cations (red), and anions (cyan), where $A^{\rm c}=A^{\rm a}=3$.}
\label{model}
\end{figure}

{\it Model and theory.} The setup consists of a periodic cubic box filled with charged spheroidal mobile ions and one (or two) fixed charged rod(s).
The simulations are carried out in the $NVT$ ensemble where we employ the Nos\'e-Hoover thermostat~\cite{nose1984, hoover1985} with a coupling constant of $0.2$ ps and reference temperature $T=300$~K.
The equations of motion are integrated using a velocity Verlet algorithm with a time step of $2$ fs to satisfy energy conservation. The production run lasts $20$ ns, out of which the first $5$ ns are omitted in the data analysis (equilibration). 
The initial configurations are prepared using Moltemplate~\cite{jewett2021} and Packmol~\cite{packmol}.
Figure~\ref{model} shows the schematic representation of our model.

The LAMMPS software was used for the simulations~\cite{plimpton1995,Thomson2022}. The interactions between components $i$ and $j$ (also between PE and ions) at a distance $r$ are modeled via a soft repulsive version of the orientation-dependent Gay-Berne potential~\cite{gay1981, brown2009}, which is obtained by shifting and truncating the potential as
\begin{equation}
      U^{ij}(\hat{\vect{u}}^{i}, \hat{\vect{u}}^{j}, {\vect{r}}^{ij})= {\epsilon}^{ij}(\hat{\vect{u}}^{i}, \hat{\vect{u}}^{j}, \hat{\vect{r}}^{ij})\left[ 4 (\Sigma_{ij}^{12}-\Sigma_{ij}^{6})+1 \right]
  \end{equation}
at $r^{ij}<r^{ij}_{\rm c} (\hat{\vect{u}}^{i}, \hat{\vect{u}}^{j},\hat{\vect{r}}^{ij})$, where
\begin{equation}
    \Sigma_{ij}=\frac{\sigma_0^{ij}}{r^{ij}-\sigma^{ij}(\hat{\vect{u}}^{i}, \hat{\vect{u}}^{j}, \hat{\vect{r}}^{ij})+\sigma_0^{ij}}.
\end{equation}
Here $\hat{\vect{u}}^{i}$ and $\hat{\vect{u}}^{j}$ are the unit vectors along the molecular axes, $\sigma^{ij}_0$ the minimum contact distance for the $ij$ pair, $\sigma^{ij}$ the orientation-dependent separation distance at which attractive and repulsive contributions cancel, ${\epsilon}^{ij}=\epsilon_{0}^{ij}[\epsilon^{ij}(\hat{\vect{u}}^{i}, \hat{\vect{u}}^{j})]^{\rm \nu}[\tilde{\epsilon}^{ij}(\hat{\vect{u}}^{i}, \hat{\vect{u}}^{j}, \hat{\vect{r}}^{ij})]^{\rm \mu}$ the orientation-dependent well depth, and $r^{ij}_{\rm c} (\hat{\vect{u}}^{i}, \hat{\vect{u}}^{j}, \hat{\vect{r}}^{ij})$ the position of the potential minimum (see Supplementary Material (SM)).
Following Ref.~\onlinecite{bates1999} we set $\nu=1$ and $\mu=2$.

We set $\sigma_{0}^{i}=0.4$ nm (common hydrated diameter of ions) and $\epsilon_{0}^{i}=0.1$ kcal mol$^{-1}$. The major ($\sigma_{\rm maj}$) and minor ($\sigma_{\rm min}$) axes define the aspect ratio $A = \sigma_{\rm maj}/\sigma_{\rm min}$. We set $\sigma_{\rm maj}=\sigma_{0}^{i} A^{2/3}$ and $\sigma_{\rm min} = \sigma_{0}^{i} A^{-1/3}$, which, regardless of $A$, provides a volume equivalent to the one of a sphere ($A=1$) with diameter equal to $\sigma_{0}^{i}$.
Here we consider $A = 1,2,3,$ and $4$.
The choice of $\epsilon_{\rm maj/min}^{i}$ follows Ref.~\onlinecite{everaers2003}.
\begin{figure}[b!]
\centering
  \includegraphics[width=0.95\linewidth]{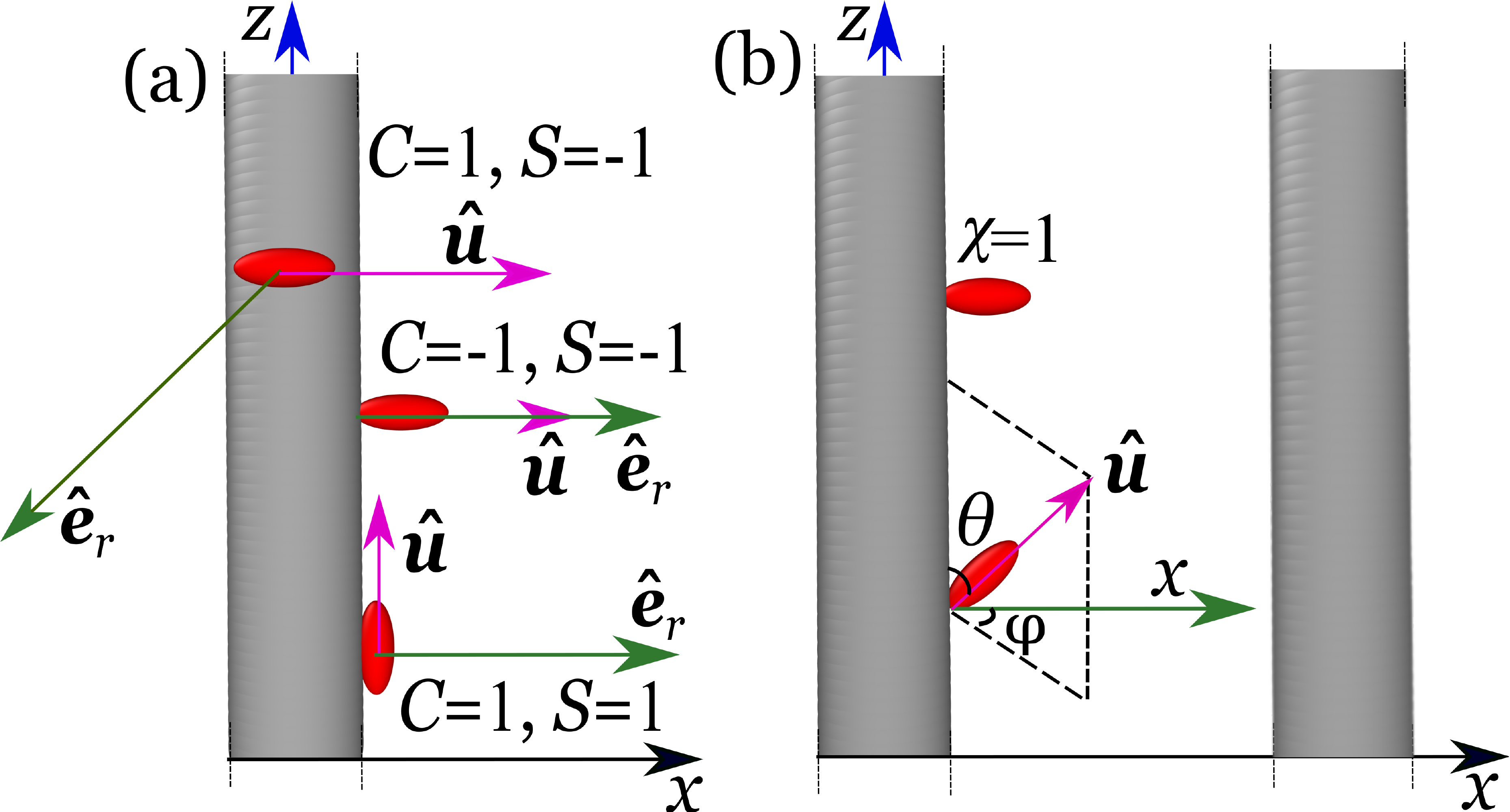}
\caption{(a) Illustration of $\hat{\vect{e}}_{r}$, $\hat{\vect{u}}$. $\hat{\vect{e}}_{r}$ is normal to the cylinder surface, whereas $\hat{\vect{u}}$ points along the ion major axis. $\hat{\vect{u}}$ defines the orientation of the ion and characterizes the relative orientation with respect to the PE via $ \hat{{\vect{e}}}_{r}\cdot \hat{{\vect{u}}}$.
Configurations relative to the PE surface and their orientation order parameters $C$ and $S$. (b) Order parameter $\chi$ is defined between two PEs. $\chi=1$ if the ion is parallel to the $x$ axis and $\chi=-1$ if perpendicular.}
\label{order_param}
\end{figure}

The rigid PE is built of charged spherical beads (force centers) interacting via the Weeks-Chandlers-Andersen~\cite{anderson} potential $U^{\rm PE}(r)=4\epsilon_{\rm PE}[(\sigma_{\rm PE}/r)^{12}-(\sigma_{\rm PE}/r)^6]+\epsilon_{\rm PE}$ for $r\leq2^{1/6}\sigma_{\rm PE}$.
Here, the bead diameter is $\sigma_{\rm PE}=1.2$ nm, and the depth of the potential well $\epsilon_{\rm PE}$ is equal to $\epsilon_0^i$. The beads are fixed at a distance $b = 0.27$ nm apart so that a smooth equipotential surface is experienced by the ions. 
The PE dimensions are in line with common synthetic and biopolymers, such as poly(styrene sulfonate) (PSS). 
Lorentz-Berthelot mixing rules $\epsilon^{ij}_{0}=\sqrt{\epsilon^{i}_{0}\epsilon^{j}_{0}} $ and $\sigma^{ij}_0 =(\sigma^{i}_0+\sigma^{j}_0)/2$ are used. 
We use $74$ beads, each with charge $Z^{\rm B} e = -e$, providing a line charge density $\lambda = Z^{\rm B} e/b \approx -4$ e/nm, close to that of PSS ($-3.7$ e/nm). The surface charge density $\lambda/\pi\sigma_{\rm PE}$ is close to that of DNA molecules ($-1$ e/nm$^2$). 

The electrostatic interactions are modelled via Coulombic potentials, which, for two charges $Z^ie$ and $Z^je$, read $\beta e V^{ij}(r)= Z^iZ^j\ell_{\rm B}/r,$
where $\beta=1/k_{\rm B}T$, the Bjerrum length $\ell_{\rm B}=\beta e^2/(4\pi \varepsilon) =0.7$ nm measures the coupling strength by specifying the distance at which two unit charges have interaction energy of $\approx k_{\rm B}T$. In this, $\varepsilon=\varepsilon_{\rm r}\varepsilon_{0}$ is the effective dielectric constant, $\varepsilon_{\rm r}$ being the solvent dielectric constant (for water, $\varepsilon_{\rm r}=78$ at $300$ K and $1$ atm~\cite{lide2004}) and $\varepsilon_0$ the vacuum dielectric constant.

These contributions are obtained in reciprocal space, after a real space cutoff of $1.2$ nm, using the Particle-Particle Particle-Mesh summation method~\cite{plimpton1997} with relative force accuracy of $10^{-5}$.
Monovalent, divalent, and trivalent charges are equally split into two points at distances of $\sigma_{\rm maj}/4$ from the center of the ions along the major axis, as sketched in Fig.~\ref{model}.
Finally, for valency ($Z$) and aspect ratio ($A$) of cations (c) and anions (a) we use the notation $Z^{\rm c}:Z^{\rm a}$ and $(A^{\rm c}, A^{\rm a})$, respectively.  

The PE is neutralized with counterions from multivalent added salt. The case of monovalent counterions with multivalent added salt is discussed in the SM.
An extra monovalent anion is added for systems containing trivalent salt counterions.
Finite size effects are checked by repeating the simulations for boxes with sides of $4$, $6$, $10$, $20$, $40$, and $60$ nm. All density profiles converge for boxes with sides of $20$ nm.

The single-charge number density distribution of species $i$ is obtained via
\begin{equation}
    n^i(r)=\langle\sum_{k=1}^{N^i}\frac{Z^i}{2}\delta(\lvert\vect{r}_{\rm }-\vect{r}_{k}^{\,i}\lvert)\rangle_{t}/V_k(r),
\end{equation}
where $\vect{r}=(x,y)$ is the distance vector on the $xy$ plane from the center of the backbone of the PE, $N_i$ the number of charges of type $i$, $\langle...\rangle_{t}$ the time average, $\vect{r}_{k}^{\,i}=(x_{k}^{\,i}, y_{k}^{\,i})$ are planar vectors pointing on single charges and $V^k(r)$ the volume of a cylindrical shell located at $r$. 
Hereafter cation and anion charge densities are denoted as $n^{\rm +}$ and $n^{\rm -}$, respectively.

To characterize the orientation along the PE $z$ axis, we define the order parameter
$S(r) \equiv 2 \langle \mid \hat{\vect{e}}_{z} \cdot \hat{\vect{u}}_{k} \mid \rangle_{t,r}-1,$
where $\langle...\rangle_{t,r}$ is both time average and average over particles at $r$, $\hat{\vect{u}}_{\rm k}$ the unit vector along the major-axis of the $k^{\rm th}$ ion and $\hat{\vect{e}}_{z}$ the unit vector along the $z$ axis. For ions oriented perpendicular to the PE $S=-1$, parallel $S=1$, and randomly oriented $S=0$.
As an additional measure of 
the tendency of the ions to be tangential to the PE surface, we define
$
    C(r) \equiv  1-2 \langle\mid \hat{\vect{e}}_{r} \cdot \hat{\vect{u}}_{k} \mid\rangle_{t,r},
$
where $\hat{\vect{e}}_{r}$ is the unit vector normal to the PE surface. If all ions are tangent to the PE surface $C=1$, if perpendicular $C=-1$, if randomly oriented $C=0$. 
Finally, to quantify the tendency to orient along the $x$ axis we use $\chi(\eta)=2\langle\lvert\sin\theta(\eta) \cos\varphi(\eta)\lvert\rangle_{t, \eta}-1,$
where $\theta$ and $\varphi$ are defined in Fig.~\ref{order_param}(b) and $\eta=(x,y)$. Specifically, $\chi=1$ for parallel, $\chi=-1$ for perpendicular and $\chi=0$ for random orientations. Figure~\ref{order_param} shows different ion orientations and the respective values of $S$, $C$ and $\chi$.

{\it Results for a single PE.}
\begin{figure}[h!]
\centering
  \includegraphics[width=1\linewidth]{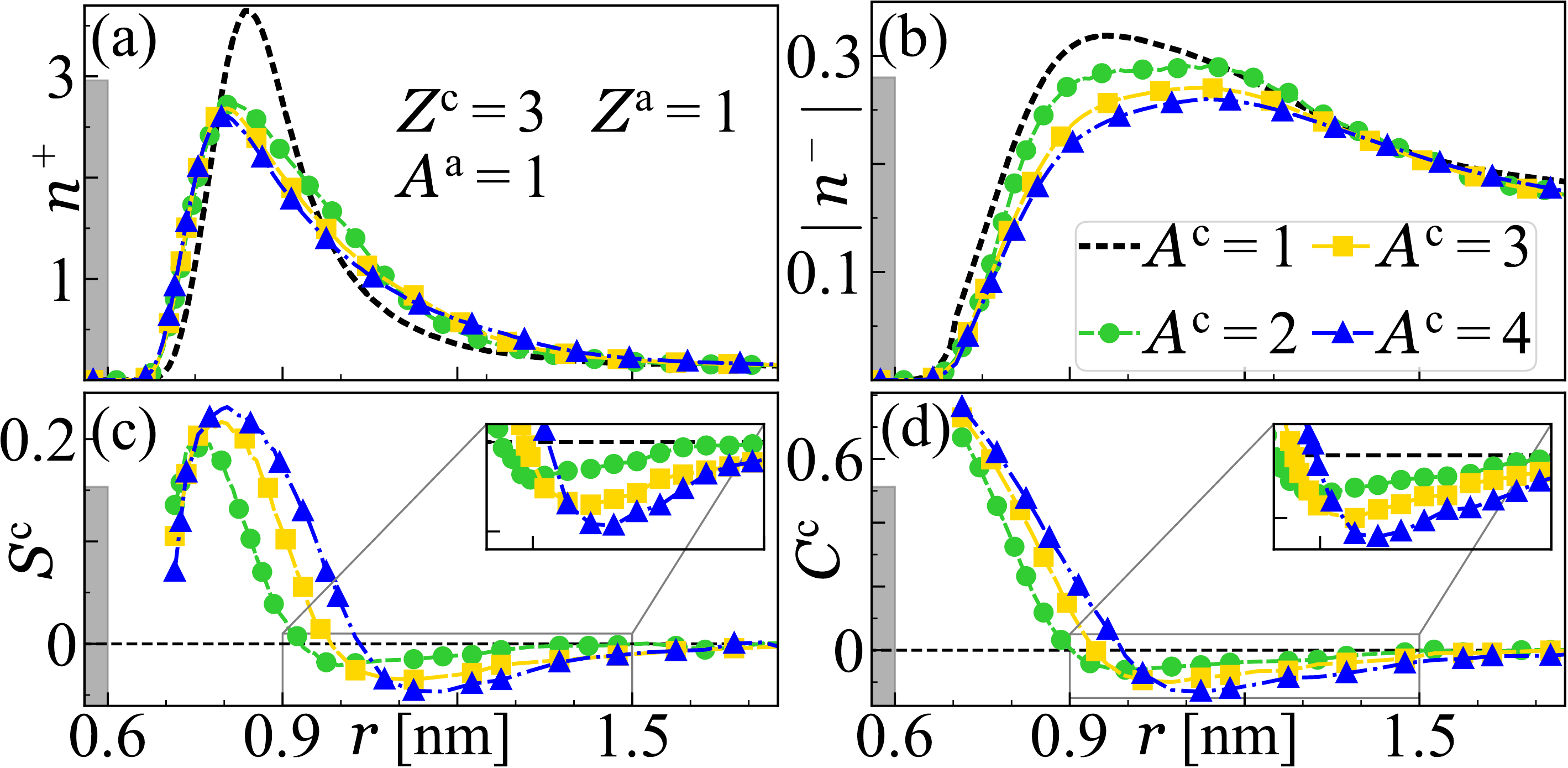}
 \caption{(a) $n^+$ and (b) $\mid n^-\mid$ ($e$ nm$^{-3}$) in $3:1$ salt with ionic strength of $0.5$ M, $A^{\rm a}=1$ and varying $A^{\rm c}$. 
(c) Order parameters $S^{\rm c}(r)$ and (d) $C^{\rm c}(r)$ for the systems shown in (a). The gray bar indicates the PE with a radius of $0.6$ nm. Error bars in this and the other figures are comparable to the symbol sizes or smaller.
}
\label{cation_3_1}
\end{figure}
We first focus on the case of ion condensation around a single PE. We have recently shown that, for monovalent salt, a soft-potential-modified PB theory gives accurate results in the case of a cylindrical PE for a wide range of salt and ion sizes~\cite{vahid2022}. When multivalent ions are introduced into the system, such mean-field approximation breaks down. To this end, we have considered three different cases at ionic strengths of $0.5$ M in detail. (i) Trivalent cations with spherical monovalent anions, i.e. case $3:1$ and $A^{\rm a}=1$ with $A^{\rm c}=1-4$. (ii) Trivalent cations and anions, i.e. case $3:3$ with $A^{\rm a}=3$ and $A^{\rm c} = 2,3,4$. (iii) Trivalent cations and anions ($3:3$) with $A^{\rm c}=3$ and $A^{\rm a} = 2,3,4$. Additional data for the effect of the ionic strength $I$, $Z^{\rm c}$, and $\lambda$ are shown in Figs.~S2, S3, and S4 of SM, respectively.  

\begin{figure}[thb]
\centering
  \includegraphics[width=1\linewidth]{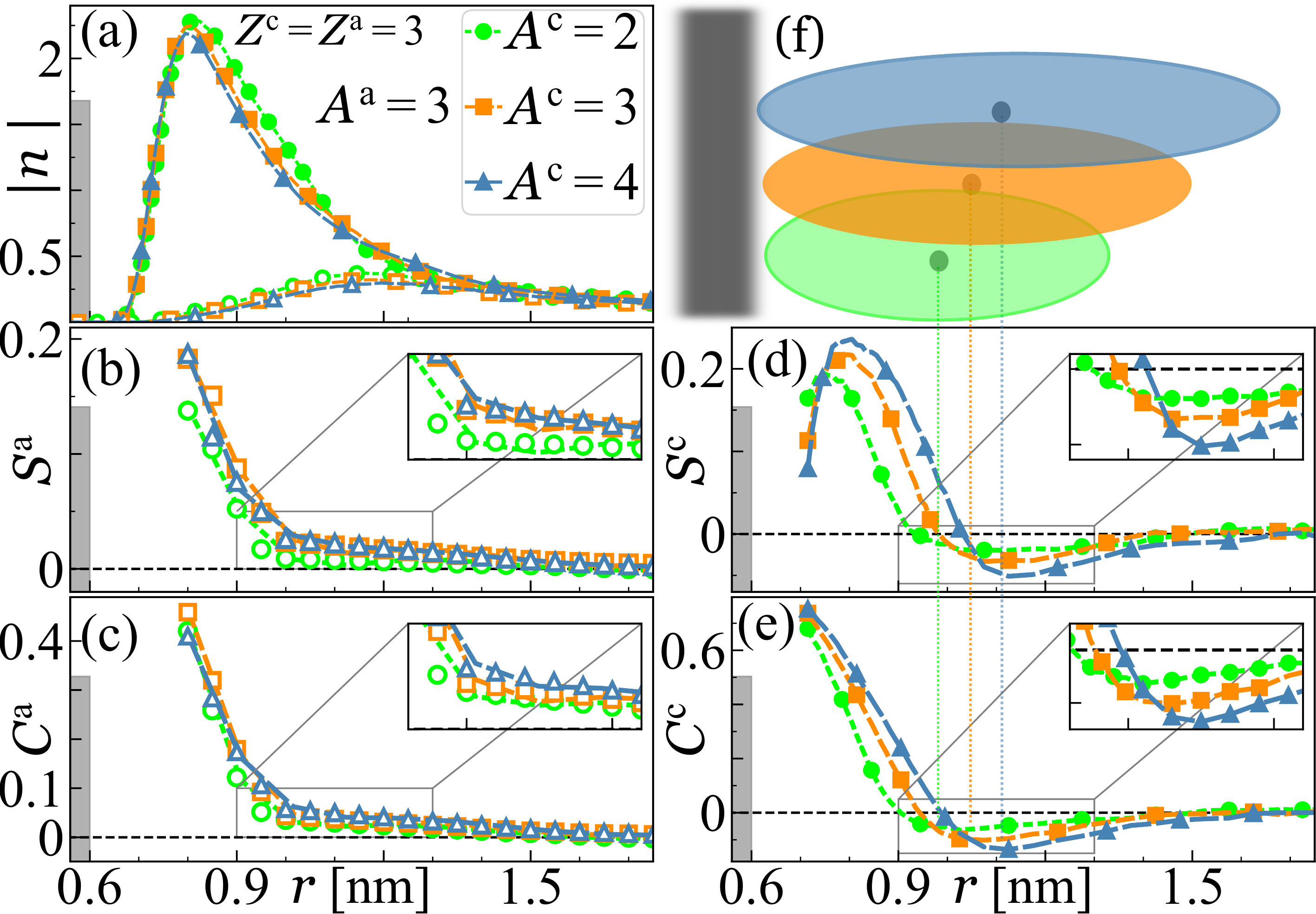}
 \caption{(a) $\mid n^+\mid$ (solid symbols) and $\mid n^-\mid$ (open symbols) ($e$/nm$^3$) in $3:3$ salt with ionic strength of $0.5$ M, $A^{\rm a}=3$ and varying $A^{\rm c}$. Order parameters
(b) $C^{\rm a}(r)$, (c) $S^{\rm a}(r)$,
(d)  $C^{\rm c}(r)$ and (e) $S^{\rm c}(r)$ for the systems in (a).
The gray bar indicates the PE with a radius of $0.6$ nm.
(f) Schematic of cations with different anisotropy oriented perpendicular to the PE, showing that the negative minima are close to $\sigma_{\rm maj}/2$.}
\label{cation_3_3}
\end{figure}

The strong electrostatic attraction between the multivalent cations and the PE results in overcharging, as shown in Refs.~\onlinecite{nguyen2000, grosberg2002, hsiao2006} and a 
large peak in $n^+$. The excess charge attracts anions, resulting in the formation of a second layer. This can be clearly seen in 
Figs.~\ref{cation_3_1}(a)-(b) and \ref{cation_3_3}(a)-(b), where we show results for the cases (i) and (ii), respectively. Results for (iii) are very similar to (ii) and are shown in Fig.~S6.
 
Interestingly, the $n^+$ data of the case (i) shows the aspherical cations have a much lower density near the PE surface than the spherical ones. This is in line with Monte Carlo simulations for dumbbell-like ions~\cite{kim2008, bohinc2012}. Furthermore, increasing the spacing between charges in the cations decreases the charge density close to PE (cf. Fig.~S7).

The position of the spheroidal cation peaks indicates orientational ordering near the PE. To quantify this, the two order parameters are shown in 
Figs.~\ref{cation_3_1}(c)-(d) and
\ref{cation_3_3}(c)-(d). They indicate that the spheroids have a tendency to align along the backbone of the PE, as expected from electrostatics. Such ordering is enhanced with increasing electrostatic interactions, as shown in Fig. S4. Curiously, the order parameters show small negative minima indicating a tendency to align perpendicular to the PE. For case (ii), the closest spheroidal anions show a tendency to align with the cations whilst exhibiting no minimum. The positions of the order-parameter minima for cations correspond to $r\approx (\sigma_{\rm PE}+\sigma_{\rm maj})/2$, as shown in Fig.~\ref{cation_3_3}(f).

\begin{figure}[bht!]
\centering
  \includegraphics[width=1\linewidth]{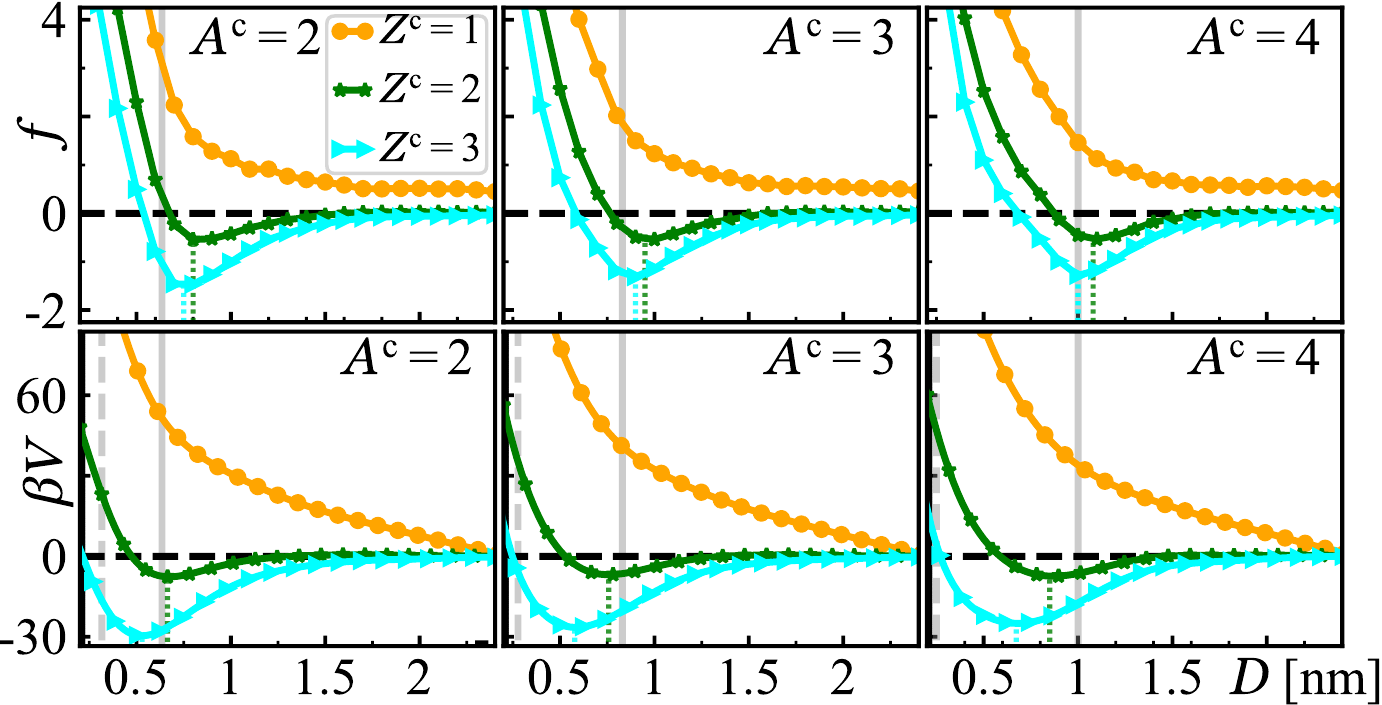}
 \caption{Normalized mean force $f(D)$ (kcal mol$^{-1}$ nm$^{-2}$) and $\beta V(D)$ for $A^{\rm c}=2, 3,$ and 4 at $Z^{\rm c}=1,2,$ and 3. Error bars are comparable to symbol sizes. 
Gray solid and dashed lines indicate $\sigma_{\rm maj}$ and $\sigma_{\rm min}$ for the cations, respectively. Dotted vertical lines indicate the positions of the minima for the curves.}
\label{2polymer_forces_only}
\end{figure}

{\it Results for PE-PE interactions.} We now turn to the interesting question of how multivalent spheroidal ions influence PE-PE interactions.
In Ref.~\onlinecite{yang2022}, we investigated the interactions between two negatively charged rods in monovalent spherical salt solutions, where the interactions are always repulsive for small ions when there are no correlations.
Various strong-coupling approaches and Monte Carlo simulations have shown that charge correlations often lead to effective charge reversal of the PE and attractive interactions between them \cite{jimenez2006,gronbech1997, shklovskii1999, naji2004, kanduvc2009, cha2018}.
To quantify the PE-PE interactions, we have computed the mean force $f(D)=F(D)/L$ ($L$ being the PE length) and the corresponding potential $V(D) = \int_D^\infty F(x')dx'$ as a function of the surface-to-surface PE distance $D$. Here $F(D)$ is the time-averaged total force acting between the PEs over production runs of $10$ ns for discrete values of $D$. In Fig.~\ref{2polymer_forces_only} we show the results for spheroids with $A^{\rm c}=2-4$. As expected, the interactions for monovalent cations remain repulsive, and the cation shape and size have only a small influence~\cite{yang2022}.  

The situation drastically changes when multivalent counterions are introduced. The forces for divalent and trivalent cations for different shapes are shown with green and cyan dots in Fig.~\ref{2polymer_forces_only}. Here both charge correlations and ion shapes play an important role. A deep negative minimum in the interaction potential appears, indicating strong binding. The values of the binding energy can be found in Table S1. The first notable result is that the minimum in $f(D)$ approaches $D \approx \sigma_{\rm maj}$ with increasing anisotropy and valency. Second, the equilibrium position corresponding to the potential minimum settles at $\approx (\sigma_{\rm min}+\sigma_{\rm maj})/2$.

\begin{figure}[thb!]
\centering
  \includegraphics[width=1\linewidth]{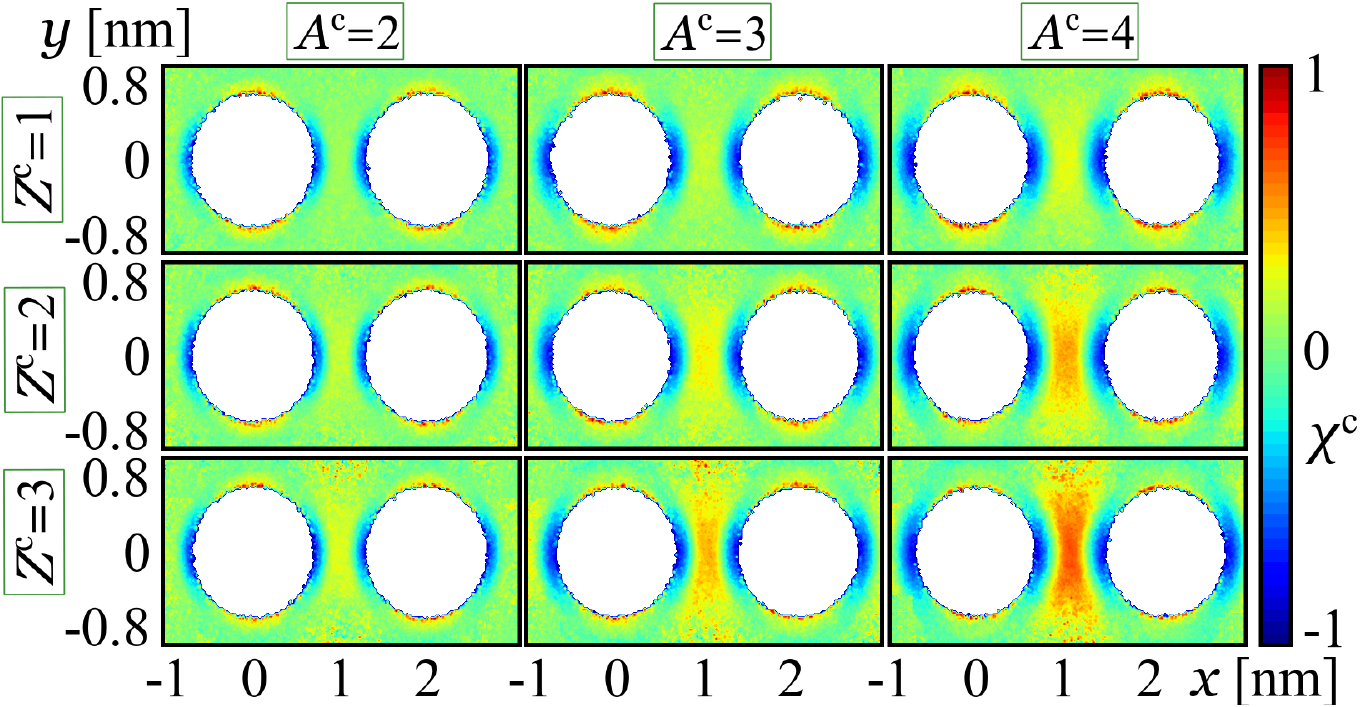}
 \caption{Contour plots of $\chi^{\rm c}$ at the maximum attractive force (for $Z^{\rm c}=1$ the distance from $Z^{\rm c}=2$ is used). Contour plots at the equilibrium distance are shown in Fig.~S8.}
\label{2polymer_counter}
\end{figure}

In Fig.~\ref{2polymer_counter}, we show contour plots of the order parameter $\chi^{\rm c}$ at values of $D$ where the attractive force is largest. They reveal an interesting correlation to the cation orientation. 
Similar to the single-PE case, cations close to the PEs tend to align tangentially, whereas, in the region between the two PEs, parallel to the $x$ axis ($\chi>0$). The middle and right columns of Fig.~\ref{2polymer_counter} show that the cations bridge the PEs at the minimum $D\approx \sigma_{\rm maj}$, thus mediating attraction between the PEs~\cite{urbanija2008, podgornik2004, maset2009, cha2018}. 
The attractive region gradually moves toward higher values of $D$ as $A^{\rm c}$ is increased since longer spheroids need more space to fit between the PEs. Larger values of $Z^{\rm c}$ lead to stronger electrostatic interactions, thus stronger attraction, but also to slightly smaller optimal distances between the PEs.

The corresponding contour plots at the equilibrium distance $f=0$ are shown in Fig. S8. Interestingly, the orientational ordering of the cations between the PEs diminishes with increasing valency. This clearly demonstrates the role of orientation in mediating the forces. At the equilibrium distance, the cation-mediated attraction and the PE-PE repulsion exactly cancel out, and cations with higher valency require less ordering to neutralize the repulsion.

To study the influence of salt, we added monovalent $1:1$ salt ($A^{\rm c}=A^{\rm a}=3$) 
to the aforementioned systems, and the results are shown in Fig.~S10.
The attraction between PEs reduces because electrostatic interactions between them weaken. An interesting opposite case is that of monovalent counterions with multivalent salt in Fig.~S11. The trivalent cations immediately lead to an attraction as charge correlations build up. The attraction is only weakly affected by added $3:1$ salt up to 1 M. 

{\it Summary and Conclusions.}
In this Letter, we have shown how in addition to valency, both ion shape and concentration can be harnessed as tools to control like-charged PE-PE interactions. Multivalent ions induce an attractive force \cite{gronbech1997, shklovskii1999, naji2004, kanduvc2009, cha2018} whose magnitude and range can be tuned by the characteristics of the ions. In particular, we found a direct correlation between the orientational ordering of aspherical multivalent ions and the mediation of the attraction. Our results can be used as general guidelines to tune self-assembly by varying the ion properties.

{\it Acknowledgements:}
This work was supported by the Academy of Finland through its Centres of Excellence Programme (2022-2029, LIBER) under project no. 346111 (M.S.) and Academy of Finland project No. 307806 (PolyDyna) (T.A-N.). The work was also supported by Technology Industries of Finland Centennial Foundation TT2020 grant (T.A-N.).
We are grateful for the support by FinnCERES Materials Bioeconomy Ecosystem. Computational resources by CSC IT Centre for Finland and RAMI -- RawMatters Finland Infrastructure are also gratefully acknowledged.
\bibliography{references}

 \pagebreak
 
\onecolumngrid
\begin{center}
{\Large \bf Supplementary Material}
\end{center}
\vspace{8mm}
\twocolumngrid

\setcounter{equation}{0}
\setcounter{table}{0}
\setcounter{figure}{0}

\renewcommand{\theequation}{S\arabic{equation}}
\let\oldthetable\thetable
\renewcommand{\thetable}{S\oldthetable}
\renewcommand{\thefigure}{S\arabic{figure}}

\section{soft repulsive Gay-Berne potential }
The original form of the GB potential for anisotropic particles reads as~\cite{gay1981}
\begin{equation}
      U_{\rm GB}^{ij}(\hat{\vect{u}}^{i}, \hat{\vect{u}}^{j}, {\vect{r}}^{ij})= 4{\epsilon}^{ij}(\hat{\vect{u}}^{i}, \hat{\vect{u}}^{j}, \hat{\vect{r}}^{ij})(\Sigma_{ij}^{12}-\Sigma_{ij}^{6}),
  \end{equation}
where
\begin{equation}
    \Sigma_{ij}=
    \frac{\sigma_0^{ij}}{r^{ij}-\sigma^{ij}(\hat{\vect{u}}^{i}, \hat{\vect{u}}^{j}, \hat{\vect{r}}^{ij})+\sigma_0^{ij}},
\end{equation}
$\hat{\vect{u}}^{i}$ and $\hat{\vect{u}}^{j}$ being unit vectors along the molecular axes, $\sigma^{ij}_0$ the minimum contact distance for $ij$ pair, $\sigma^{ij}$ the orientation-dependent separation distance at which attractive and repulsive contributions cancel, and ${\epsilon}^{ij}=\epsilon_{0}^{ij}[\epsilon^{ij}(\hat{\vect{u}}^{i}, \hat{\vect{u}}^{j})]^{\rm \nu}[\tilde{\epsilon}^{ij}(\hat{\vect{u}}^{i}, \hat{\vect{u}}^{j}, \hat{\vect{r}}^{ij})]^{\rm \mu}$ the orientation-dependent well depth. 

\begin{figure}[bht]
\centering
  \includegraphics[width=0.8\linewidth]{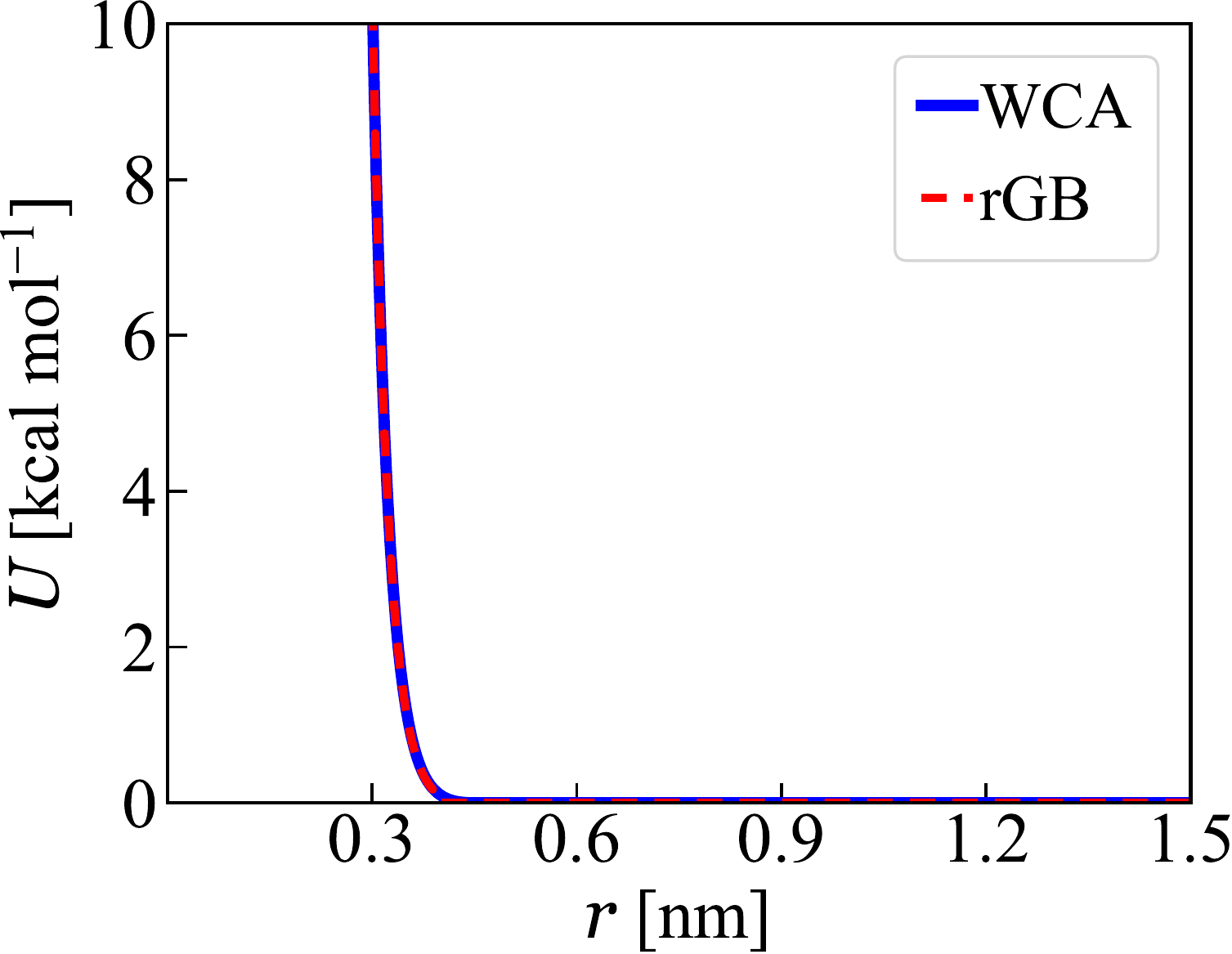}
 \caption{Comparing the potential energy $U(r)$ between two neutral beads calculated using rGB and WCA potentials in LAMMPS. The effective diameter of the particles is $0.4$ nm. }
\label{rgb_pot}
\end{figure}

This potential is already implemented in the LAMMPS simulation package~\cite{plimpton1995,Thomson2022, brown2009}, but in our study, the polymer beads and ions interact with each other via a soft repulsive GB potential (rGB), which is not implemented in LAMMPS. This potential is obtained by shifting and truncating the GB potential:
\begin{equation}
    U^{ij}=\begin{cases}
    U_{\rm GB}^{ij}(\hat{\vect{u}}^{i}, \hat{\vect{u}}^{j}, {\vect{r}}^{ij})+{\epsilon}^{ij}(\hat{\vect{u}}^{i}, \hat{\vect{u}}^{j}, \hat{\vect{r}}^{ij})
    , \>  r^{ij}<r^{ij}_{\rm c}; \\
    0, \> {\rm otherwise};\end{cases}
\end{equation}
where $r^{ij}_{\rm c} (\hat{\vect{u}}^{i}, \hat{\vect{u}}^{j}, \hat{\vect{r}}^{ij})=(2^{1/6}-1)\sigma^{ij}_{0}+\sigma^{ij}(\hat{\vect{u}}^{i}, \hat{\vect{u}}^{j}, \hat{\vect{r}}^{ij})$ is the position of the potential minimum. 
To add this potential to LAMMPS, we modified the original potential file and it is provided at \href{https://ida.fairdata.fi/s/NOT_FOR_PUBLICATION_c2LfbNEWyk25}{\url{https://ida.fairdata.fi/s/NOT\_FOR\_PUBLICATION\_c2LfbNEWyk25}} [permanent link is available in published manuscript].
 Please cite the main article DOI: \{this paper\} if the potential or its modifications are used in your work.
 We tested this potential using the 3Mar2020 stable version of LAMMPS to verify that it functions properly. Figure~\ref{rgb_pot} shows a test example results for two spheres with a diameter of $0.4$ nm and $\epsilon^{ij}_{0}=0.1$ kcal mol$^{-1}$. The interaction between them is calculated using the rGB potential in LAMMPS, which in this case is expected to reduce to the isotropic Weeks-Chandlers-Andersen (WCA) potential~\cite{anderson}:
 \begin{equation}
   U_{\rm WCA}(r)=4\epsilon_{\rm}[(\sigma_{\rm}/r)^{12}-(\sigma_{\rm}/r)^6]+\epsilon_{\rm},   
 \end{equation}
with $\sigma=0.4$ nm and $\epsilon=0.1$ kcal mol$^{-1}$.

\section{Additional results for a single PE}
We present additional results of MD simulations for the single PE case. We investigate the effect of salt concentration, cation valency, PE line charge density $\lambda$, anion shape, spacing between charges within ions, charge distribution, as well as ion orientation and ordering. In all cases, the PE diameter, ion shapes, box size, and computational details are the same as those described in the main text unless otherwise stated.

\subsection{Effect of ionic strength}
\begin{figure}[bht]
\centering
  \includegraphics[width=1\linewidth]{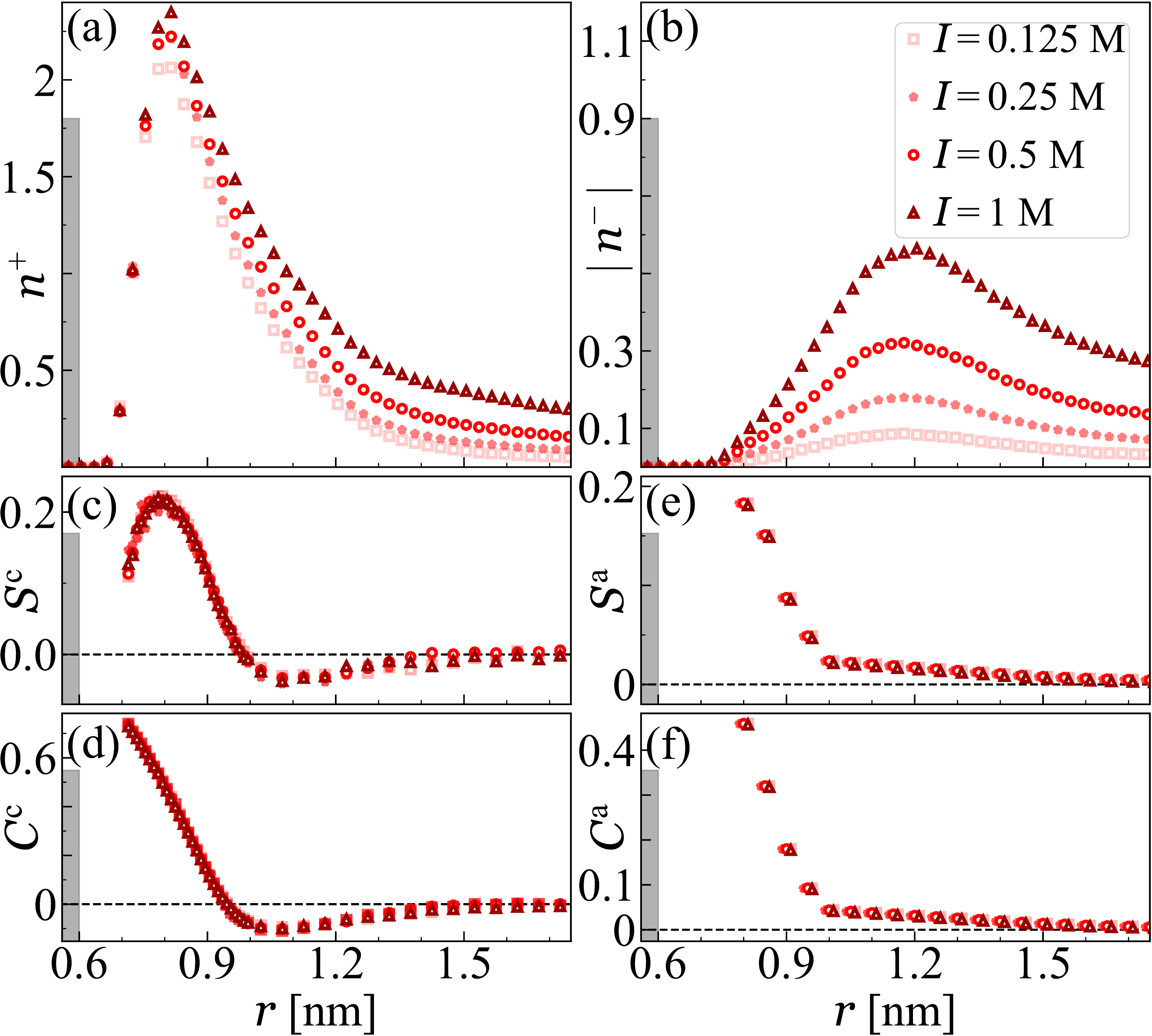}
 \caption{(a) $\mid n^+\mid$ and (b) $\mid n^-\mid$ ($e$/nm$^3$) in $3:3$ salt with $A^{\rm c}=A^{\rm a}=3$ for different ionic strengths. Order parameters
(c) $S^{\rm c}(r)$, (d) $C^{\rm c}(r)$,
(e)  $S^{\rm a}(r)$ and (f) $C^{\rm a}(r)$ for the systems in (a).
The gray bar indicates the PE with a radius of $0.6$ nm. Error bars in this and the other figures are comparable to symbol sizes or smaller.}
\label{salt_effect}
\end{figure}

We have performed additional MD simulations to characterize the effect of ionic strength on ion condensation and ordering.
We carried out the simulations for case (ii) at ionic strength $I$ ranging from $0.125$~M to $1$~M. In these systems, both cations and anions are trivalent ($3:3$) and possess the same aspect ratio of $A^{\rm c}=A^{\rm a}=3$.
Figure~\ref{salt_effect} shows the effect of $I$ on the charge distribution and orientation ordering of the ions. The results show that increasing $I$ leads to an increase in $n^+$ and $\mid n^-\mid$, as expected. An interesting result is that the orientation of the ions is insensitive to salt which.   

\subsection{Effect of cation valency}
\begin{figure}[hbt]
\centering
  \includegraphics[width=1\linewidth]{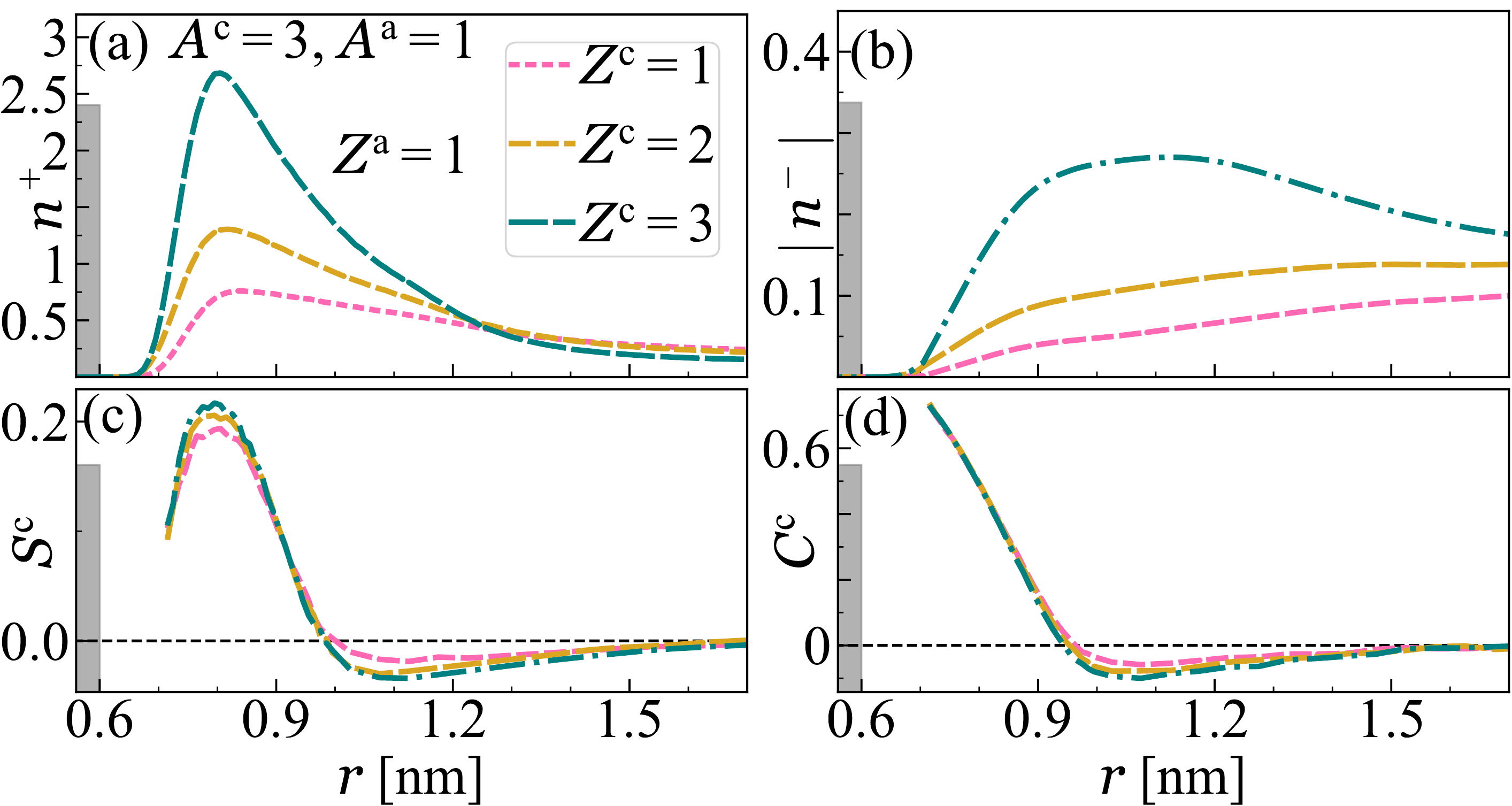}
 \caption{(a) $n^+$ and (b) $\mid n^-\mid$ ($e$/nm$^3$) in $3:1$ salt with $I=0.5$ M, $A^{\rm c}=3$, $A^{\rm a}=1$ and varying $Z^{\rm c}$. Order parameters
(c) $S^{\rm c}(r)$, (d) $C^{\rm c}(r)$ for the systems in (a).
The gray bar indicates the PE with a radius of $0.6$ nm.
}
\label{valency-effect}
\end{figure}

We have further studied the effect of cation valency on charge density profiles and cation orientational ordering. 
We have performed additional simulation for case (i) at $Z^{\rm a}=1$ and different cation valency $Z^{\rm c}$. 
Figure~\ref{valency-effect} shows the effect of $Z^{\rm c}$ on $n^+$, $\mid n^-\mid$, $S^{\rm c}$, and $C^{\rm c}(r)$ in $Z^{\rm c} = 1$ salt for $Z^{\rm c}=1,2,3$ where $I=0.5$ M. Cations are aspherical with $A^{\rm c}=3$, and anions are monovalent and spherical. Panels (a) and (b) show that the higher the cation valency, the higher the peaks in $n^+$ and $\mid n^-\mid$. Panel (c) shows that the orientational ordering of the cations slightly increases with increasing  $Z^{\rm c}$.

\subsection{Effect of linear charge density of a single PE}
\begin{figure}[hbt]
\centering
  \includegraphics[width=1\linewidth]{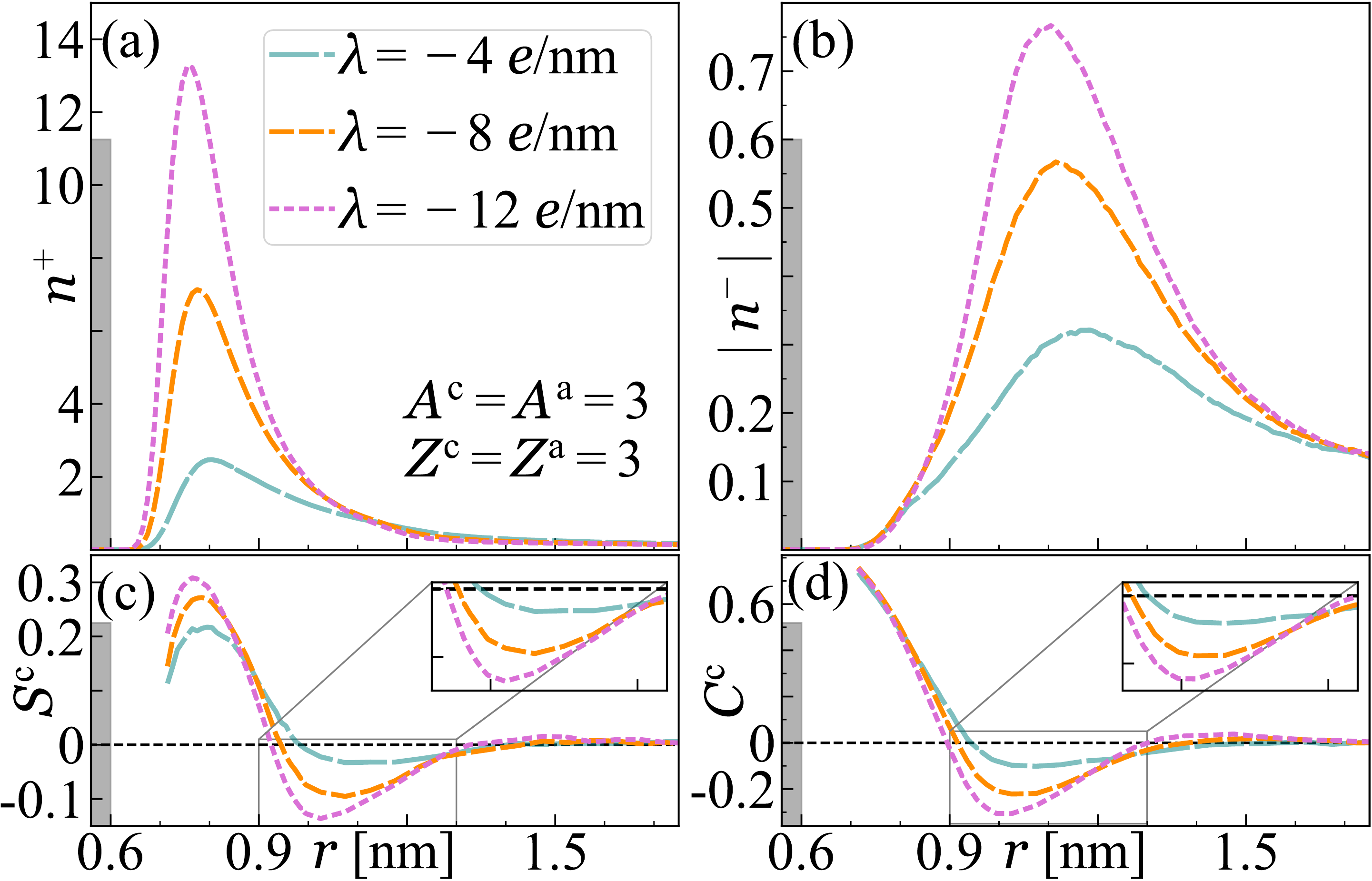}
 \caption{(a) $n^+$ and (b) $\mid n^-\mid$ ($e$ nm$^{-3}$) in $3:3$ salt with ionic strength of $0.5$ M, $A^{\rm c}=A^{\rm a}=3$, and varying $\lambda$. 
 The cation order parameter (c)  $S^{\rm c}(r)$ and (d) $C^{\rm c}(r)$ for the systems in (a).
}
\label{singlePE_charge}
\end{figure}

In this section, we demonstrate the effect of $\lambda$ on $n^+$, $\mid n^-\mid$, $S^{\rm c}$, and $C^{\rm c}$ for case (ii) at $A^{\rm c}=A^{\rm a}=3$, $Z^{\rm c}=Z^{\rm a}=3$, and $I=0.5$ M. 
In Fig.~\ref{singlePE_charge}, we show that $n^+$ and $\mid n^-\mid$ increase significantly close to the PE backbone as $\lambda$ is increased. This is accompanied by a tendency to align along the backbone of the PE as a function of $\lambda$, as shown in panels (c) and (d). This is due to an increase in the electrostatic interactions close to the PE. Additionally, the tendency to align perpendicular to the PE at larger distances is also enhanced as indicated by the negative minimum.  

\subsection{Effect of anion aspect ratio}

To understand the effect of anion shape on charge distributions and orientational ordering, we carried out additional simulations for different anion aspect ratios $A^{\rm a}$.
To this end, we considered two different cases. The first one is that of trivalent cations with monovalent anions, i.e. the case $3:1$ and $A^{\rm c}=1$ with $A^{\rm a}=2,3,4$. The second case has trivalent cations and anions ($3:3$) with $A^{\rm c}=3$ and $A^{\rm a} = 2,3,4$, which is the case (iii) in the main text. 

\subsubsection{3:1 salt}
\begin{figure}[hbt]
\centering
  \includegraphics[width=1\linewidth]{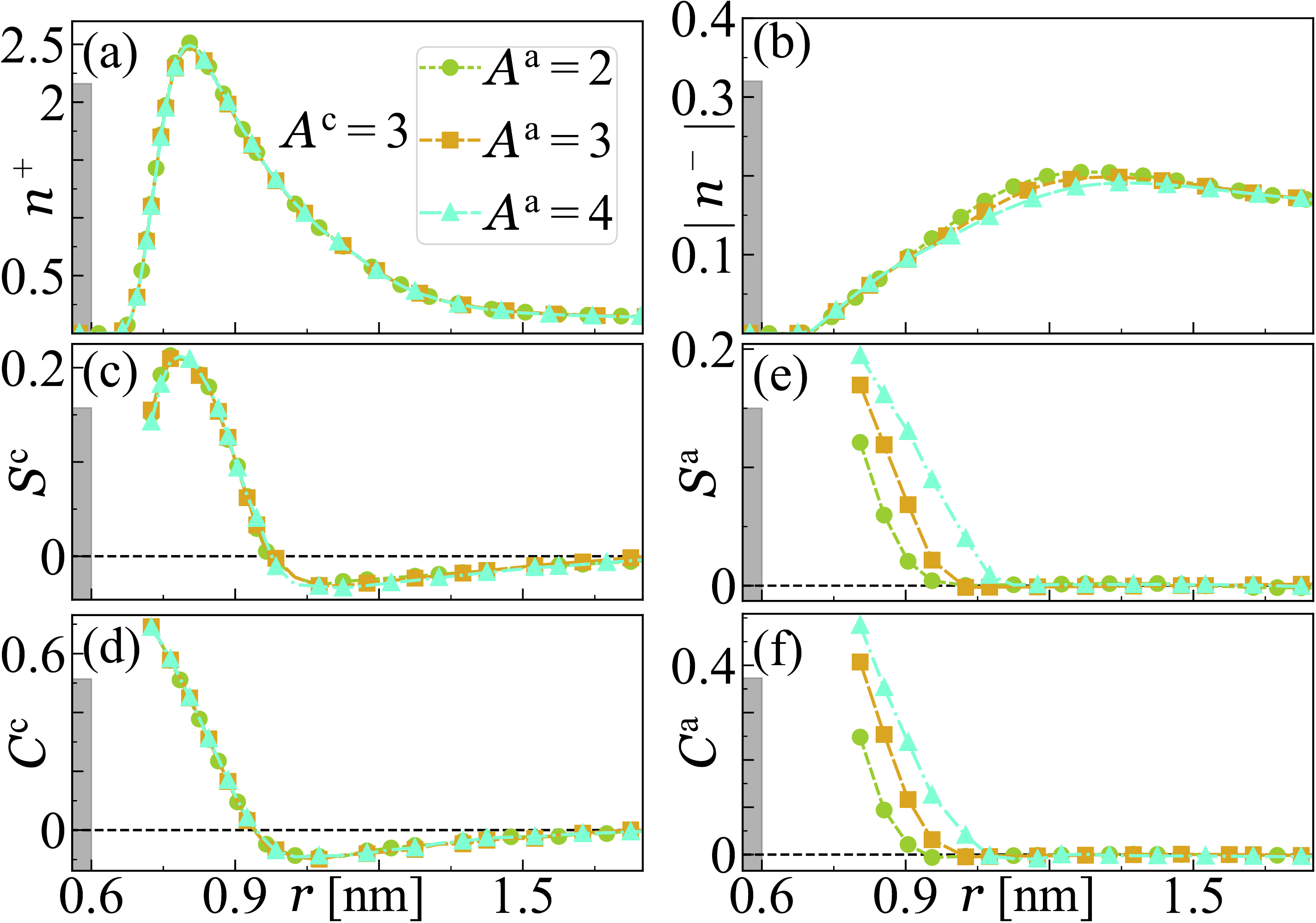}
 \caption{(a) $\mid n^+\mid$ and (b) $\mid n^-\mid$ ($e$/nm$^3$) in $3:1$ salt with ionic strength of $0.5$ M, $A^{\rm c}=3$ and varying $A^{\rm a}$. Order parameters
(c) $S^{\rm c}(r)$, (d) $C^{\rm c}(r)$,
(e)  $S^{\rm a}(r)$ and (f) $C^{\rm a}(r)$ for the systems in (a).
The gray bar indicates the PE with a radius of $0.6$ nm.
}
\label{anion_3_1}
\end{figure}

In Fig.~3 of the main text, we showed how $A^{\rm c}$ affects $n^+$, $\mid n^-\mid$, $S^{\rm c}$, and $C^{\rm c}$ in case (i). We observed that the shape of cations shifts the orientational order parameters. 
In this section, we consider systems similar to case (i), but instead of varying $A^{\rm c}$, we fix it at $3$ and vary $A^{\rm a}$. The ion valencies $Z^{\rm c}=Z^{\rm a}=3$ and the value of $I$ is $0.5$ M in all cases. 
In Fig.~\ref{anion_3_1}, we observe that $n^{\rm +}$, $S^{\rm c}$, and $C^{\rm c}$ are not affected by the shape of the anions. In contrast, $S^{\rm a}$ and $C^{\rm a}$ increase close to the PE backbone as $A^{\rm a}$ is increased.

\subsubsection{3:3 salt}
\begin{figure}[tbh]
\centering
  \includegraphics[width=1\linewidth]{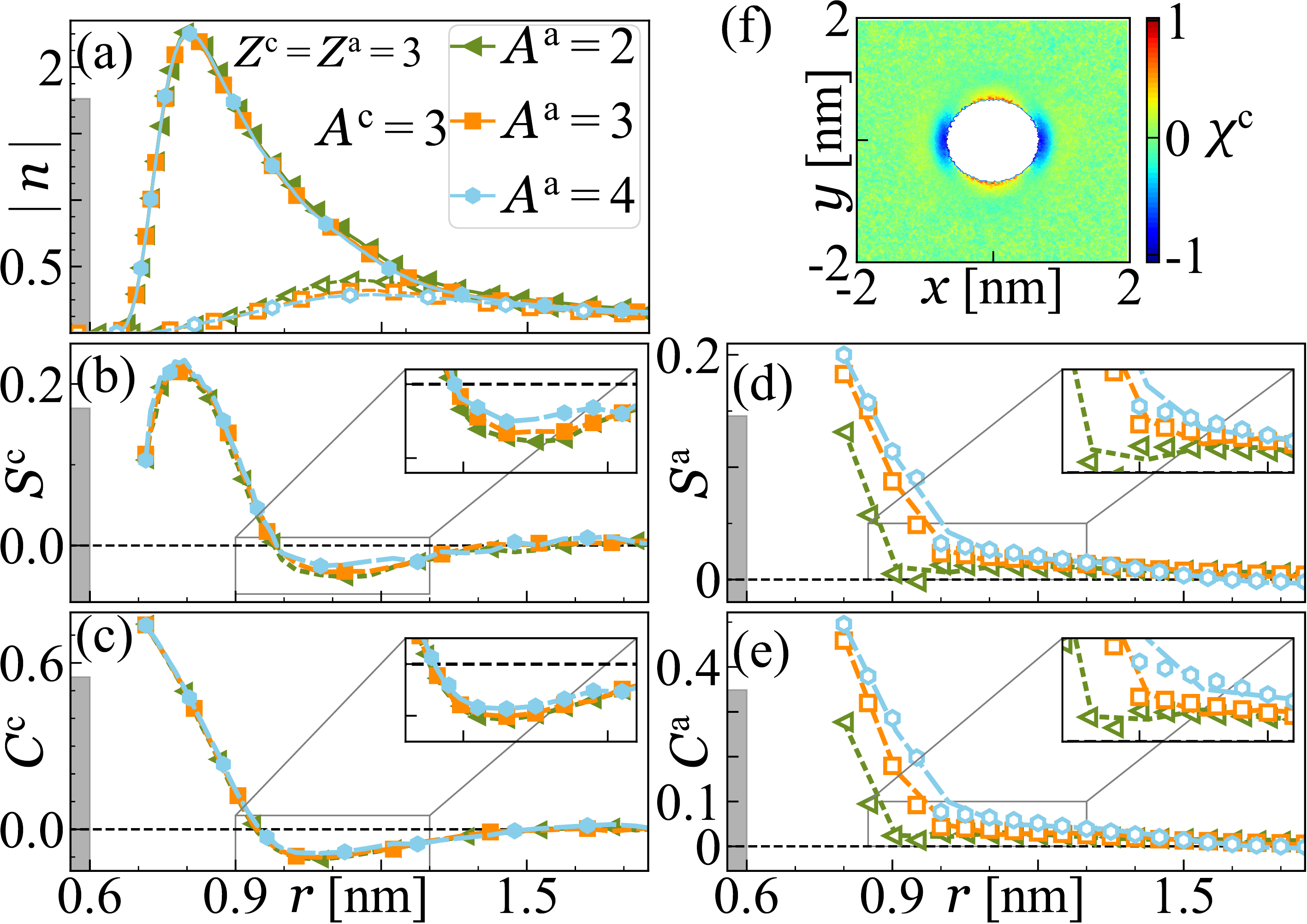}
 \caption{(a) $\mid n^+\mid$ (solid symbols) and $\mid n^-\mid$ (open symbols) ($e$/nm$^3$) in $3:3$ salt with ionic strength of $0.5$ M, $A^{\rm c}=3$ and varying $A^{\rm a}$. Order parameters
(b) $S^{\rm c}(r)$, (c) $C^{\rm c}(r)$,
(d)  $S^{\rm a}(r)$ and (e) $C^{\rm a}(r)$ for the systems in (a).
The gray bar indicates the PE with a radius of $0.6$ nm.
(f) Contour plot of $\chi^{\rm c}$. 
 }
\label{anion_3_3}
\end{figure}

Here, we show additional data for case (iii). 
The main difference between cases (ii) and (iii) in the main text is that in case (ii),  $A^{\rm a}$ is fixed at $3$ and $A^{\rm c}$ is varied, whereas in case (iii), $A^{\rm c}$ is fixed at $3$ and $A^{\rm a}$ is varied.
In Fig.~\ref{anion_3_3}, we vary $A^{\rm a}$ from $2$ to $4$ at $Z^{\rm c}=Z^{\rm a}=3$ and $A^{\rm c}=3$. The values of $S^{\rm a}$ and $C^{\rm a}$ near the PE backbone are larger for longer ions. Meanwhile, practically no difference can be perceived in $n^+$, $S^{\rm c}$ or $C^{\rm c}$. In panel (f), we show a two-dimensional plot of $\chi^{\rm c}$, used as a comparison with the two-PE case. The order parameter
$\chi^{\rm c}$ quantifies the tendency to orient along the $x$ axis (which for the two-PE case becomes the axis connecting the centers of the PEs).
Because $\chi^{\rm c}$ is not radially invariant, it shows different values in different directions near the PE here.
It decays to zero quickly close to the PE, indicating random orientation along the $x$ axis. 

\subsection{The effect of spacing between charges within the cations}
\begin{figure}[hbt]
\centering
  \includegraphics[width=1\linewidth ]{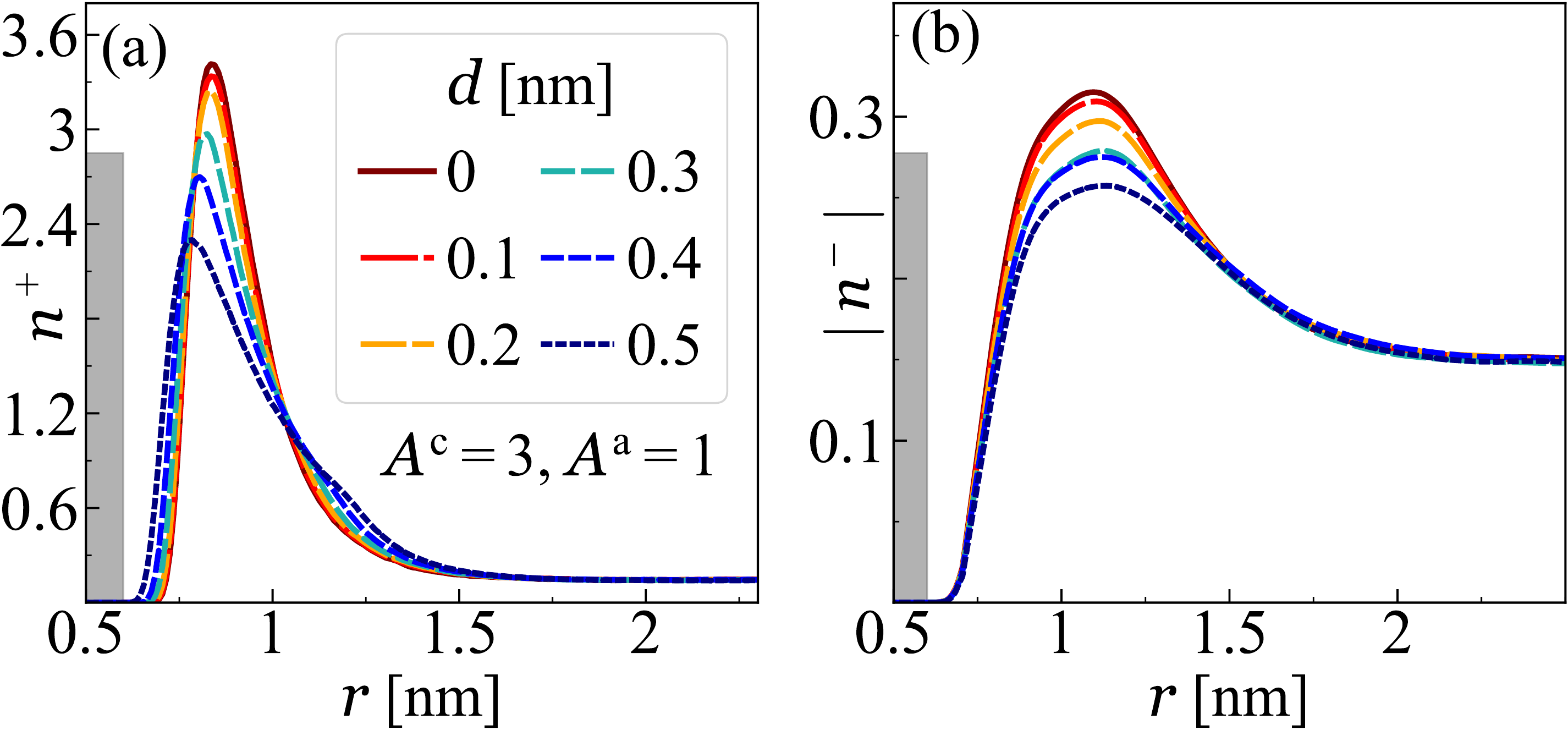}
  \caption{(a) $n^+$ and (b) $\mid n^-\mid$ ($e$/nm$^3$) in $3:1$ salt with ionic strength of $0.5$ M, $A^{\rm c}=3$ and $A^{\rm a}=1$ for different spacings between charges within the cations.}
\label{spacing_effect}
\end{figure}

In Fig.~3 of the main text, we observed that aspherical cations have a much lower density near the PE backbone than the spherical ones. One of the reasons for this behavior is the spacing between charges $d$ in aspherical cations.
In this section, we show how varying $d$ leads to differences in charge distributions.
We have conducted additional simulations for the case (i) at $I=0.5$ M where $Z^{\rm c}=3$, $Z^{\rm a}=1$, $A^{\rm c}=3$, $A^{\rm a}=1$, varying $d$ from $0$ to $0.5$ nm.   
Figure~\ref{spacing_effect} shows the effect of spacing between charges on $n^+$ and $\mid n^-\mid$.
Increasing the spacing between the charges within the cations leads to a decrease in the peaks of $n^+$ and $\mid n^-\mid$, and the cations are pushed further away from the PE backbone. 

\section{Additional results for PE-PE interactions}

Here, we first summarize additional results for binding energy and maximum attractive force for different valencies and shapes. Then, we show the contour plots of $\chi^{\rm c}$ and $S^{\rm c}$ at equilibrium distances and the distances where the attractive forces are largest, respectively.
Finally, additional data to study the effect of added salt on PE-PE interactions are included. 

\subsection{The values of binding energy and maximum attractive force for PE-PE interaction}

The binding energy $E_{\rm B}$ is defined as the amount of energy required to separate two PEs from their equilibrium distance to infinity, and can be measured from the minimum of the potential of the mean force $V(D)$.
In Table~\ref{tab1}, we present the binding energy and maximum attractive force for varying $Z^{\rm c}$ and $A^{\rm c}$. The data shows that the binding energy is greater for shorter and more charged cations than for longer and less charged ones. The distance
$D^{f}_{\rm m}$ is defined as the minimum of $f(D)$, and $D^{V}_{\rm m}$ is where $V(D)$ is minimum. 
We note that $D^{f}_{\rm m}$ approaches $D \approx \sigma_{\rm maj}$ and $D^{V}_{\rm m}$ approaches $\sigma_{\rm a}=(\sigma_{\rm maj}+\sigma_{\rm min})/2$ with increasing anisotropy and valency. 

\begin{table}[t!]
\caption{ The major diameter of counterions $\sigma_{\rm maj}$ (nm), the distance where $f(D)$ is minimum $D_{\rm m}^{f}$ (nm), the maximum attractive force $f(D_{\rm m}^{f})$ (kcal mol$^{-1}$ nm$^{-2}$), $\sigma_{\rm a}$, the equilibrium distance $D^{V}_{\rm m}$ (nm), and the dimensionless binding energy $\beta E_{\rm B}$ for $A^{\rm c}=2,3$ and $4$ at $Z^{\rm c}=2$ and $3$.}
\begin{ruledtabular}
\begin{tabular}{lccccccr}
\textrm{$Z^{\rm c}$}&
\textrm{$A^{\rm c}$}&
\textrm{$\sigma_{\rm maj}$}&
\textrm{$D^{f}_{\rm m}$}&
\textrm{$f(D^{f}_{\rm m})$} & \textrm{$\sigma_{\rm a}$} & \textrm{$D^{V}_{\rm m}$}& \textrm{$\beta E_{\rm B}$}\\
\colrule
    \multirow{3}{*}{$2$} & $2$ & $0.635$  & $0.8$ & $-0.54\pm0.01$ & $0.476$ & $0.66$ & $-8.2\pm0.4$ \\
    
    & $3$ & $0.83$ &  $1$  & $-0.52\pm0.01$ & $0.553$& $0.76$ & $-7.4\pm0.4$   \\
   
    & $4$ & $1.0$ & $1.1$  & $-0.51\pm0.01$ & $0.625$ & $0.87$ & $-7.2\pm0.3$  \\
    \\
    \multirow{3}{*}{3} & $2$ & $0.635$ & $0.8$  & $-1.46\pm0.01$ &$0.476$ & $0.53$ & $-29.6\pm0.3$  \\
    
    & $3$ & $0.83$ & $0.9$  & $-1.31\pm0.01$ & $0.553$& $0.58$ & $-26.5\pm0.5$  \\
   
    & $4$ & $1.0$ & $1.0$  & $-1.28\pm0.01$ & $0.625$ & $0.67$ & $-25.0\pm0.1$ \\
\end{tabular}
\end{ruledtabular}
\label{tab1}
\end{table}

 \begin{figure}[b!]
\centering
  \includegraphics[width=1\linewidth]{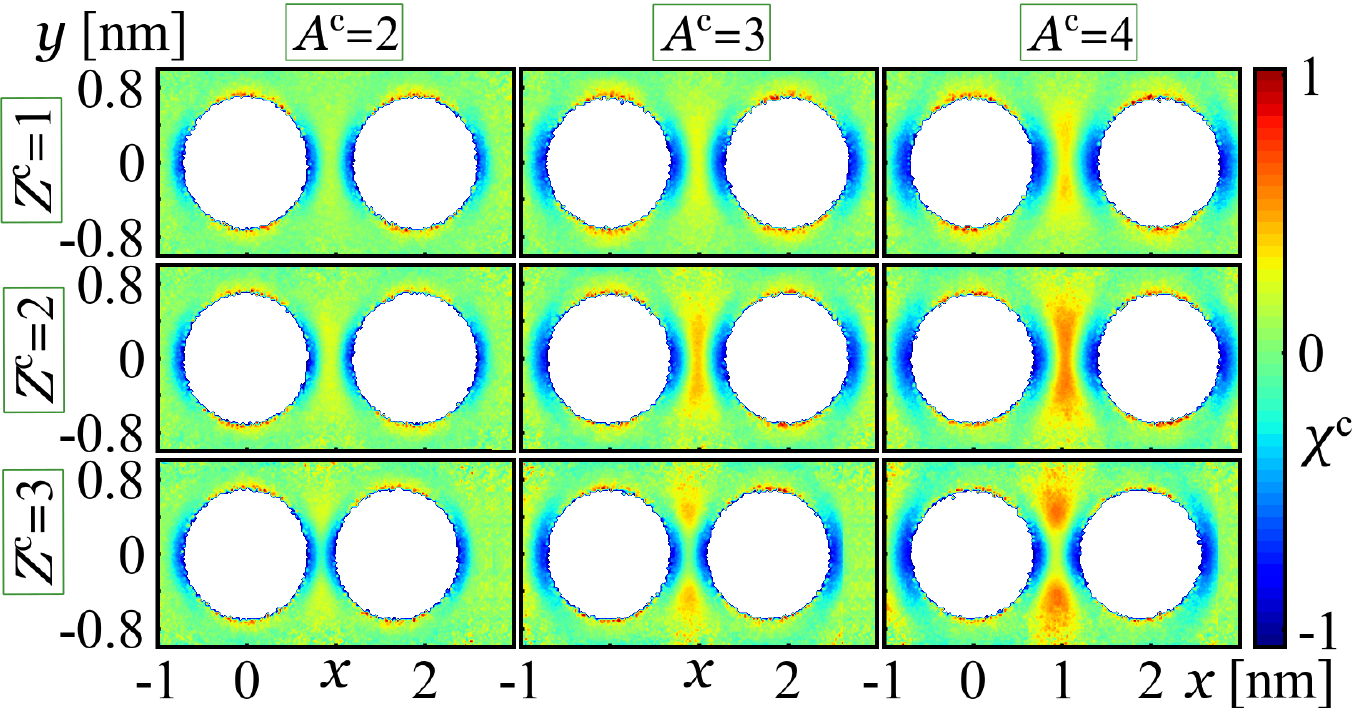}
 \caption{Contour plots of $\chi^{\rm c}$ at the equilibrium distance (for $Z^{\rm c}=1$ the distance from $Z^{\rm c}=2$ is used).}
\label{2polymer_2}
\end{figure}

\subsection{Contour plots of $\chi^{\rm c}$ at the equilibrium distance}
We provide here further analysis of counterion orientation at the equilibrium distances. 
In Fig.~\ref{2polymer_2}, we show contour plots of the order parameter $\chi^{\rm c}$ at the equilibrium distances for different counterion valencies $Z^{\rm c}$ and different counterion aspect ratios $A^{\rm c}$.
Interestingly, the orientational ordering of the cations between the PEs diminishes with increasing valency. This clearly demonstrates the role of orientation in mediating the forces. At the equilibrium distance, the cation-mediated attraction and the PE-PE repulsion exactly cancel out, and cations with higher valency require less ordering to neutralize the repulsion.

\subsection{Contour plots of $S^{\rm c}$ where the attractive force is largest}

\begin{figure}[thb!]
\centering
  \includegraphics[width=1\linewidth]{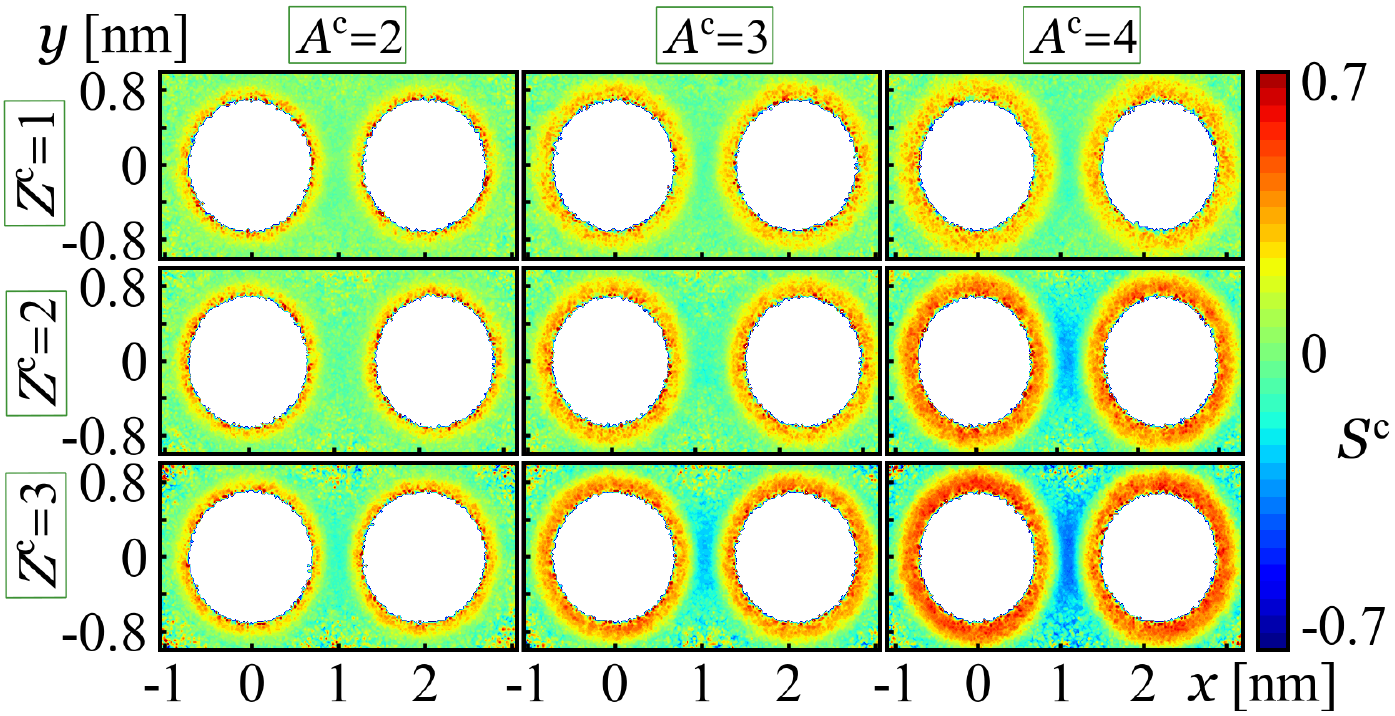}
 \caption{Contour plots of $S^{\rm c}$ at the maximum attractive force (for $Z^{\rm c}=1$ the distance from $Z^{\rm c}=2$ is used).  }
\label{2polymer_counter_zz}
\end{figure}

In Fig.~6 of the main text, we showed how the shape and valency of counterions affect their tendency to align along the $x$ axis. In this section, we study their tendency to align along the $z$ axis by means of $S^{\rm c}$.  
In Fig.~\ref{2polymer_counter_zz}, we show contour plots of the order parameter $S^{\rm c}$ at values of $D$ where the attractive force is largest.
$S^{\rm c}$ is given by $S(r) \equiv 2 \langle \mid \hat{\vect{e}}_{z} \cdot \hat{\vect{u}}_{k} \mid \rangle_{t,r}-1,$ where $\hat{\vect{u}}_{\rm k}$ the unit vector along the major axis of the $k^{\rm th}$ ion and $\hat{\vect{e}}_{z}$ the unit vector along the $z$ axis.
The contour plots show that $S^{\rm c}$ increases by increasing $A^{\rm c}$, and $S^{\rm c}$ between PEs decreases by increasing valency because the tendency to align along the $x$ axis is enhanced. 

\subsection{Effect of salt on the PE-PE interactions}

\begin{figure}[hbt!]
\centering
  \includegraphics[width=1\linewidth]{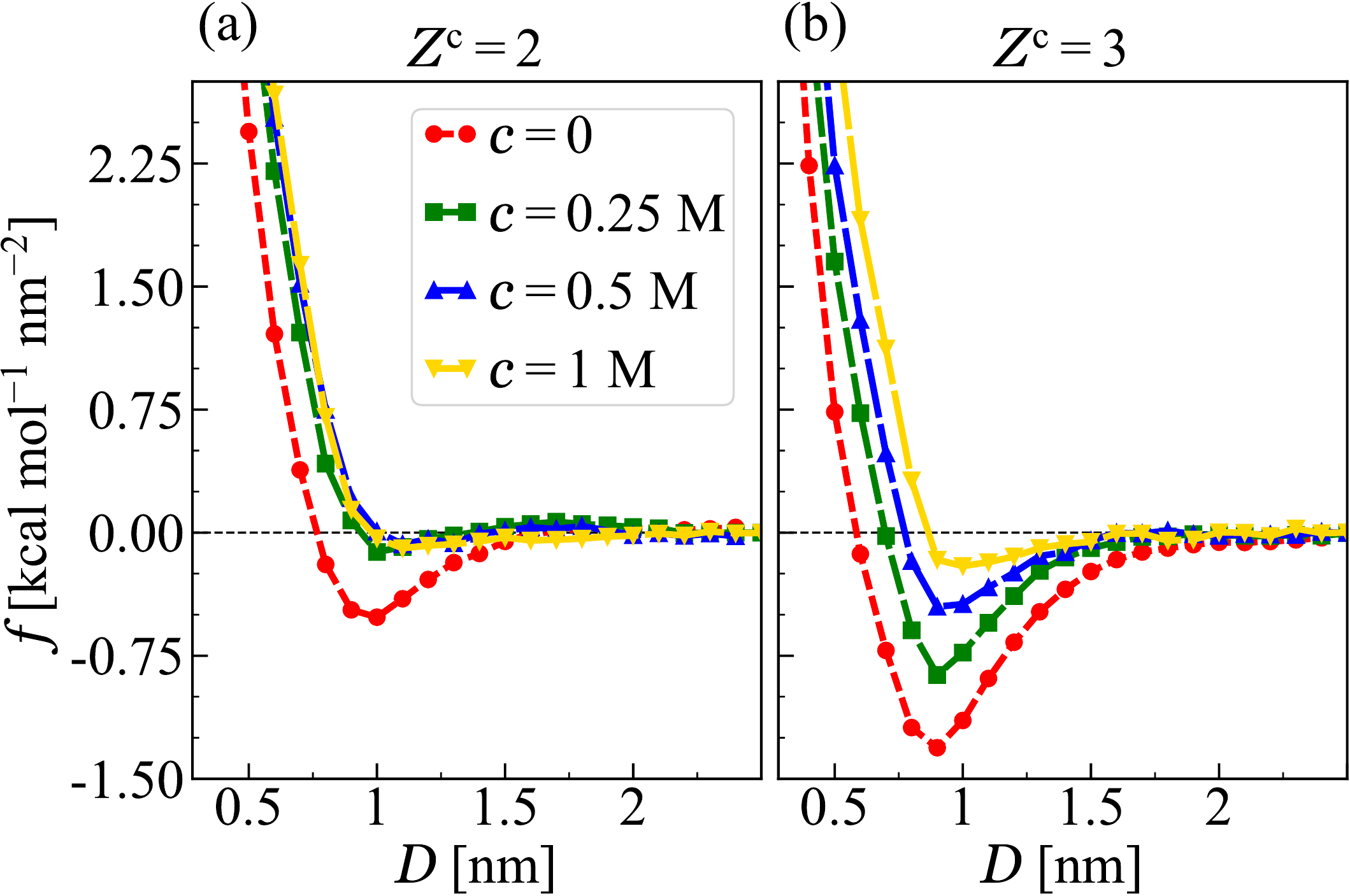}
 \caption{Normalized mean force $f(D)$ (kcal mol$^{-1}$ nm$^{-2}$) for different added monovalent salt concentrations $c_{\rm }$. In these cases, $A^{\rm c}=3$ and (a) $Z^{\rm c}=2$ and (b) $Z^{\rm c}=3$. }
\label{added_salt_2PE}
\end{figure}

\begin{figure}[hbt!]
\centering
  \includegraphics[width=0.92\linewidth]{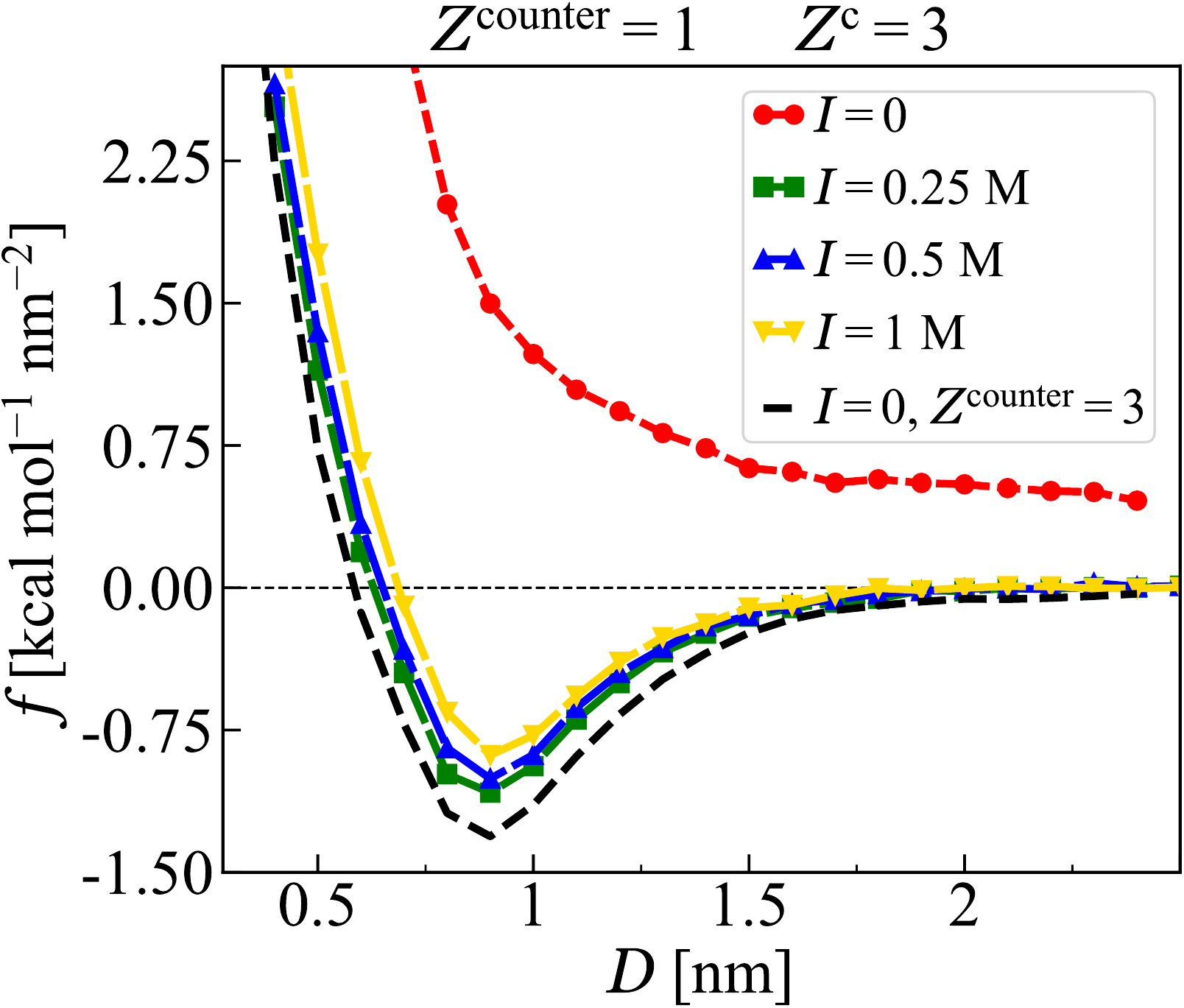}
 \caption{Normalized mean force $f(D)$ (kcal mol$^{-1}$ nm$^{-2}$) for different added $3:1$ salt with ionic strength of $I_{\rm }$. In these cases, $A^{\rm c}=3$ and counterions are monovalent. The black dashed line refers to the system with only trivalent counterions (no added salt).}
\label{added_salt_mono_counterion}
\end{figure}

In this section, two different cases are studied to explore the effect of added salt on the interaction between two PEs.
In the first case, the PE is neutralized by counterions from the added multivalent salt $Z^{\rm c}=2,3$ (cf. Fig.~\ref{added_salt_2PE}). In all systems $A^{\rm c}=A^{\rm a}=3$, whereas the salt concentration $c$ is varied.
Figure~\ref{added_salt_2PE} shows that increasing salt concentration increases the screening effect, which weakens the attraction between PEs.
In the latter case, the effect of monovalent counterions $Z^{\rm counter}=1$ with multivalent added salt $Z^{\rm c}=3$ is striking. Figure~\ref{added_salt_mono_counterion} shows that the trivalent cations immediately lead to a strong attraction as charge correlations build up. The attraction is only weakly affected by added $3:1$ salt up to 1 M.

\newpage

\end{document}